\documentclass[fleqn,usenatbib]{mnras}

\usepackage[T1]{fontenc}
\usepackage{ae,aecompl}
\usepackage{graphicx}	% Including figure files
\usepackage{amsmath}	% Advanced maths commands
\usepackage{amssymb}	% Extra maths symbols

\def\limepy{\textsc{limepy}}

\title[Double clusters in the LMC]{Three candidate double clusters in the LMC: truth or dare?}

\author[E. Dalessandro et al.]{
Emanuele Dalessandro,$^{1}$\thanks{E-mail: emanuele.dalessandro@oabo.inaf.it}
Alice Zocchi,$^{1,2}$ 
Anna Lisa Varri,$^{3}$ 
Alessio Mucciarelli,$^{1,2}$ 
\newauthor
Michele Bellazzini,$^{1}$
Francesco R. Ferraro,$^{1,2}$
Barbara Lanzoni, $^{1,2}$
Emilio Lapenna, $^{1,2}$
\newauthor
and Livia Origlia $^{1}$
\\
$^{1}$INAF - Osservatorio Astronomico di Bologna, via Gobetti 93/3, I-40129, Bologna, Italy\\
$^{2}$Dipartimento di Fisica \& Astronomia, Universit\'a degli Studi di Bologna, via Gobetti 93/2, I-40129, Bologna, Italy \\
$^{3}$Institute for Astronomy, University of Edinburgh, Royal Observatory, Blackford Hill, Edinburgh EH9 3HJ, UK \\
}

\date{Accepted XXX. Received YYY; in original form ZZZ}

\pubyear{2017}

\begin{document}
\label{firstpage}
\pagerange{\pageref{firstpage}--\pageref{lastpage}}
\maketitle

\begin{abstract}
The Large Magellanic Cloud (LMC) hosts a large number of candidate stellar cluster pairs.
Binary stellar clusters provide important clues about cluster formation processes and the evolutionary 
history of the host galaxy. However, to properly extract and interpret this information,
it is crucial to fully constrain the fraction of real binary systems and their physical properties.
Here we present a detailed photometric analysis based on ESO-FORS2 images 
of three candidate cluster multiplets in the LMC, namely SL349-SL353, SL385-SL387-NGC1922 and
NGC1836-BRHT4b-NGC1839. For each cluster we derived ages, structural parameters and morphological properties.
We have also estimated the degree of filling of their Roche lobe, as an approximate tool to measure the strength of the tidal perturbations 
induced by the LMC. 
We find that the members of the possible pairs SL349-SL353 and BRHT4b-NGC1839 have a similar age 
($t = 1.00 \pm 0.12$ Gyr and $t = 140 \pm 15$ Myr, respectively), thus possibly hinting to a common origin of their member systems
We also find that all candidate pairs in our sample show evidence of intra-cluster overdensities that can be 
a possible indication  of real binarity. Particularly interesting is the case of SL349-SL353. 
In fact, SL353 is relatively close to the condition of critical filling, 
thus suggesting that these systems might actually constitute an energetically bound pair. 
It is therefore key to pursue a detailed kinematic screening of such clusters, 
without which, at present, we do not dare making a conclusive statement about the true nature of this putative pair.

\end{abstract}


%%%%%%%%%%%%%%%%%%%%%%%%%%%%%%%%%%%%%%%%%%%%%%%%%%

%%%%%%%%%%%%%%%%% BODY OF PAPER %%%%%%%%%%%%%%%%%%

\section{Introduction}
It is commonly accepted that star clusters form from the fragmentation of giant molecular clouds in cloud cores that 
eventually produce stellar complexes, OB association or larger systems \citep{efremov1995}. 
However, the exact mechanisms of formation are not understood yet and likely there are different paths that lead to the formation of 
different systems of clusters. Particularly intriguing is the idea that star clusters could form in pairs or multiplets \citep{delafuente2009}.   

Indeed, both the Milky Way (MW) and the Large Magellanic Cloud (LMC) stellar cluster systems contain a sizable 
population ($\sim10\%$) of massive and young/intermediate age clusters, with projected mutual distance $\le 20$ pc
(\citealt{BhatiaHatz1988,Bhatia1991,Surdin1991,Subramaniam1995,Dieball2002,Mucciarelli2012,desilva2015}). 
On the other hand, in both systems double old clusters are completely lacking.  
Statistical arguments indicate that the LMC binary cluster population cannot be simply explained in terms of projection effects, 
but gravitationally bound systems should be a relevant fraction of the listed candidates \citep{BhatiaHatz1988,Dieball2002}. 
An additional handful of binary clusters are also known in the nearby Universe, in the Small Magellanic Cloud \citep[SMC;][]{HatzBhatia1990}, 
in M31 \citep{Holland1995}, in NGC5128 \citep{Minniti2004}, in the Antennae galaxies \citep{fall2005} and in the young starburst galaxy M51 
\citep{larsen2000}. 

There are three possible explanations for the origin of these systems: (1) they formed from the fragmentation of the same molecular cloud
\citep{ElmegreenElmegreen1983}, (2) they were generated in distinct molecular clouds and then became bound systems after 
a close encounter leading to a tidal capture \citep{Vallenari1998,Leon1999}, or (3) they are the result of division of a 
single star-forming region \citep{goodwin2004,arnold2017}.   
Their subsequent evolution may also have different outcomes. Dynamical models and N-body simulations \citep[see, e.g.][and references therein]{BarnesHut1986,deOliveira1998} 
have shown that, depending on the initial conditions, a bound pair of clusters may either become unbound, because of significant mass loss in 
the early phases of stellar evolution, or merge into a single and more massive cluster on a short timescale ($\approx 60$ Myr) 
due to loss of angular momentum from escaping stars \citep[see][]{PortegiesZwartRusli2007}. 
The final product of a merger may be characterized by a variable degree of kinematic and morphologic complexity, mostly depending on the values 
of the impact parameter of the pre-merger binary system \citep{deOliveira2000,Pry2016}. In some cases, the stellar system resulting from the merger event may show significant 
internal rotation 
\citep[in fact, for many years this has been the preferred dynamical route to form rotating star clusters, see][]{SugimotoMakino1989,Makino1991,Okumura1991,deOliveira1998}. 
Merger of cluster pairs has been sometimes invoked to interpret the properties of particularly massive and dynamically 
complex clusters (e.g., see the study of $\omega$ Centauri by \citealt{Lee1999}, G1 by \citealt{Baumgardt2003}, and NGC2419 
by \citealt{Bruns2011}), 
and, more in general, as an avenue to form clusters with multiple populations with different chemical abundances both in terms 
of iron and light-elements \citep[e.g.,][]{vandenBergh1996,Catelan1997,AmaroSeoane13,Gavagnin2016,Hong2017}.

The population and the properties of binary clusters could depend on the past evolutionary history of the host galaxy. 
In fact, several theoretical investigations suggest that encounters and interactions between the LMC and the SMC triggered the formation of binary systems and, 
on a larger extent, the formation of most of the known GCs in the LMC/SMC system. For example, \citet{Kumai1993} pointed out that, 
if interstellar gas clouds have large-scale random motions in the interacting LMC/SMC system, then they may collide to form compact star 
clusters through strong shock compression. \citet{Bekki2004} demonstrated that the star formation efficiency in interacting galaxies can significantly increase, 
resulting in the formation of compact stellar systems and double clusters.  This idea is supported by the link
between the two bursts of cluster formation in the LMC \citep[$\sim 100$ Myr and $1-2$ Gyr ago;][]{Girardi1995} 
and the epochs of the closest encounters between the SMC and LMC, as predicted by various theoretical models (\citealt{GardinerNoguchi1996}; 
see also \citealt{Kallivayalil2013}, for more recent models and references).

In principle, then, the study of cluster pairs provides crucial information about the mechanisms of cluster formation and evolution, 
and the possible interactions suffered by the host galaxy in the past. In practice, however, very little is known to date about these systems.

Up to now the criterion typically used to select cluster pairs has been the observed small angular separation \citep[$d<1.4\arcmin$;][]{Dieball2002} and 
the only additional hint is the evidence that in some of these candidates the two components appear to be coeval.
However, age estimates are quite uncertain, since they are usually derived exclusively from integrated colors 
\citep[e.g.,][]{Bica1996}, as rich color-magnitude diagrams (CMDs; e.g. \citealt{Vallenari1998}) are avaialble only in a few cases.
So far, the binarity has been confirmed by means of a detailed chemical analysis and
radial velocities obtained with high-resolution spectra only in the case of NGC2136 - NGC2137 in the LMC  
\citep{Mucciarelli2012} and NGC5617 - Trumpler22 in the Galaxy \citep{desilva2015}.

In this work we attempt to provide a more robust characterization of three candidate cluster pairs in the LMC: 
SL349-SL353, SL387-SL385 and NGC1836-BRHT4b.
We use three main quantities to assess their nature (i.e. possible binarity): 1) ages from Main-Sequence Turn-OFF (MSTO) luminosity in well
populated CMDs; 2) cluster structure parameters as derived by number counts of resolved stars; 3) evidence of tidal distortions and
analysis of possible signatures of interaction with their tidal environment.

The manuscript is organized as follows: Section 2 provides a description of the images obtained with FORS2 at the Very Large Telescope 
and their corresponding analysis. An age estimate of the star clusters under consideration is presented in Section 3. 
The structural and dynamical properties of the clusters are discussed in Section 4, in Section~5 we analyze tidal effects
on the stellar systems. 
Finally, we  discuss our results and present our conclusions in Section 6.

\begin{figure}
\includegraphics[width=\columnwidth]{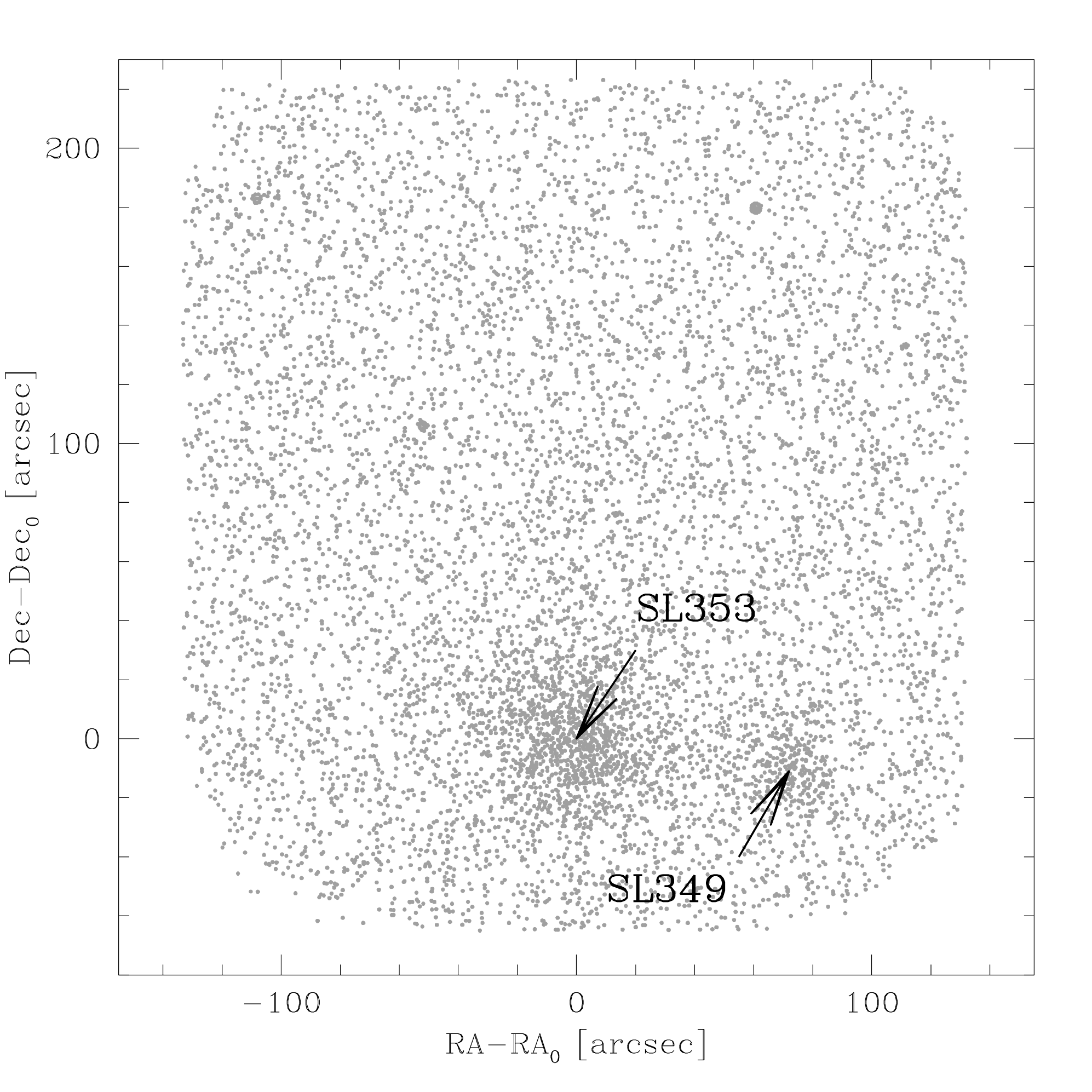}
\caption{Map of the candidate cluster pair SL349-SL353 obtained with FORS2. The centre of each cluster is indicated with an arrow.}
\label{map1}
\end{figure}

\begin{figure}
\includegraphics[width=\columnwidth]{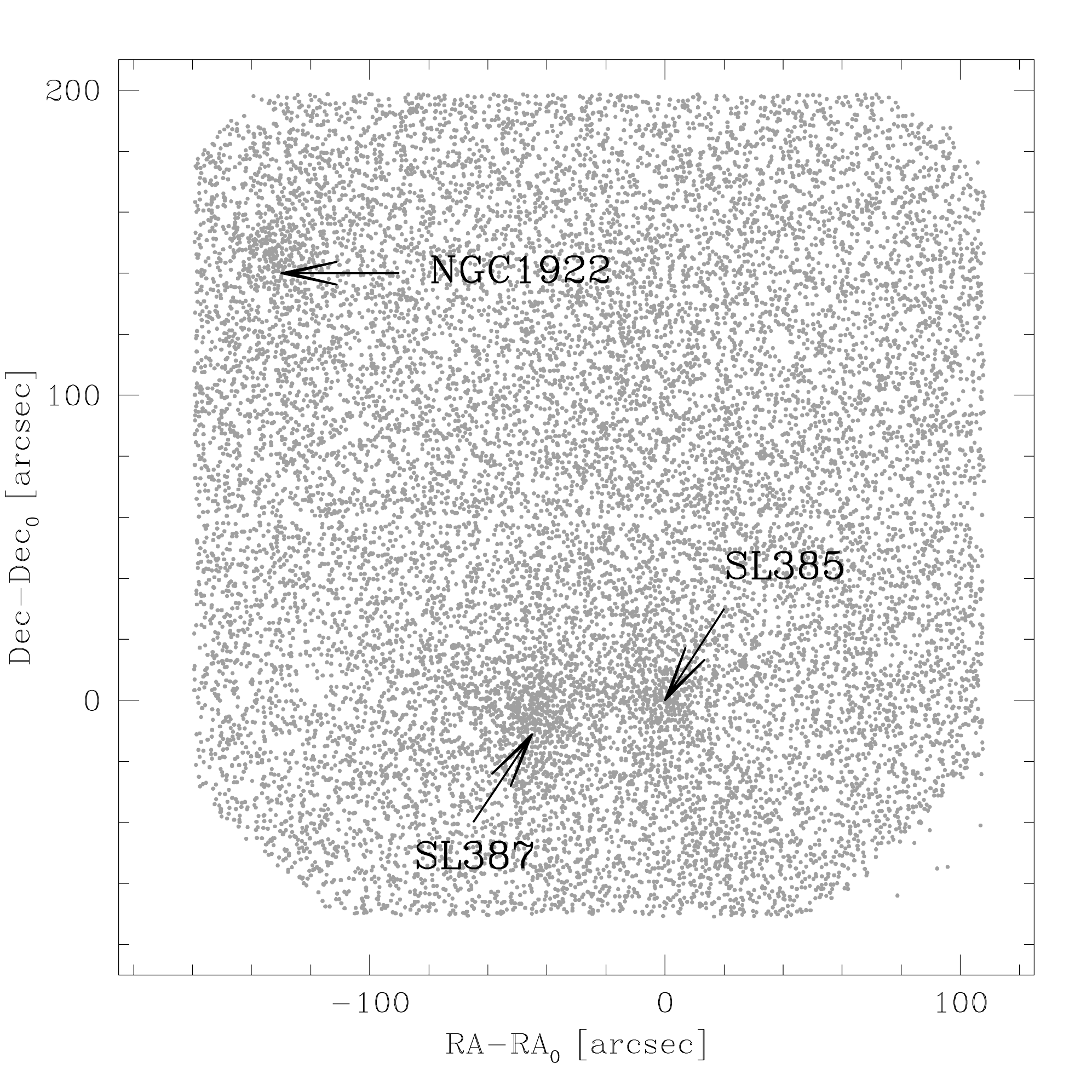}
\caption{Map of the candidate cluster pair SL387-SL385; the stellar system NGC 1922 is also visible within the field of view. The centre of each cluster is indicated with an arrow.}
\label{map2}
\end{figure}  

\section{Observations and data analysis}

\begin{figure}
\includegraphics[width=\columnwidth]{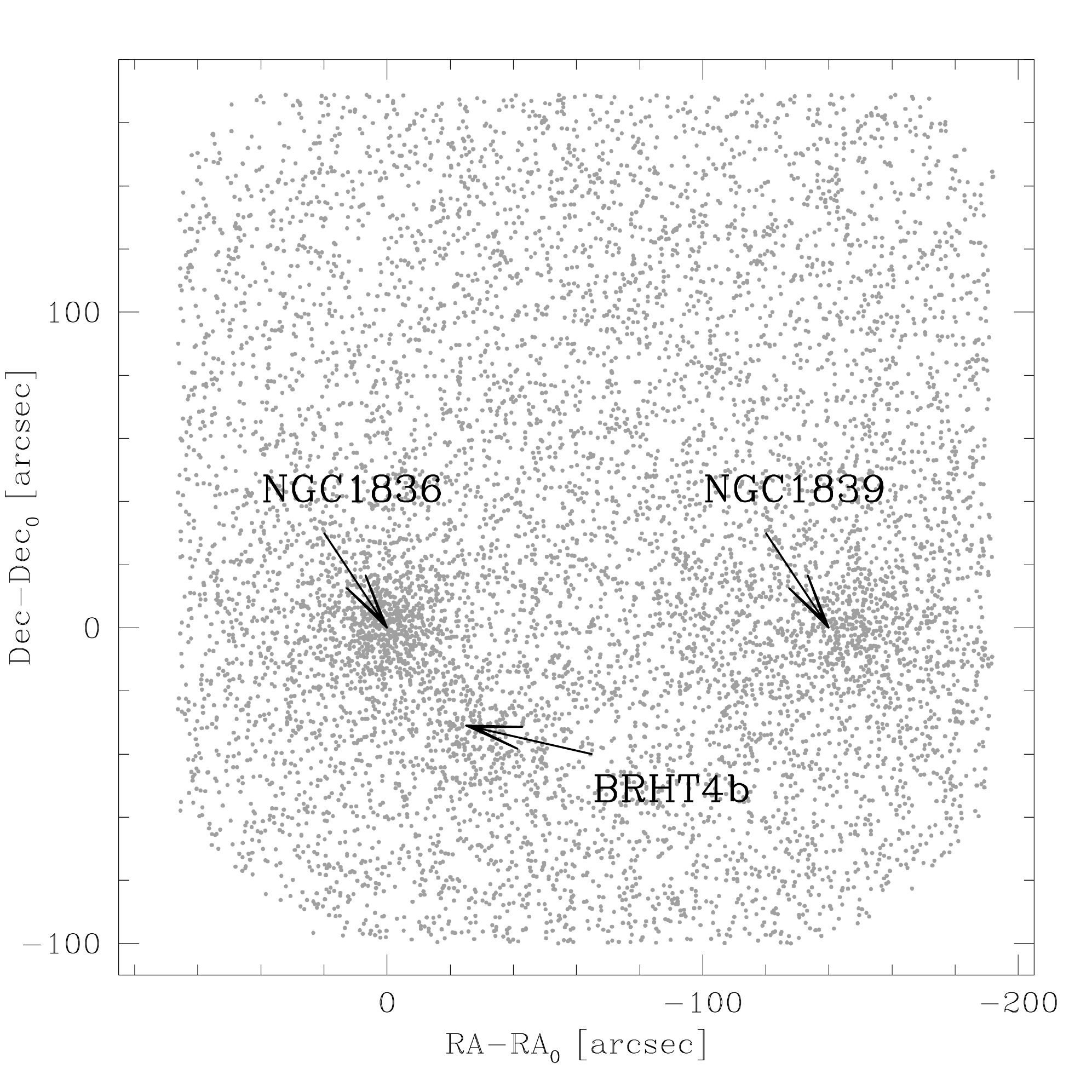}
\caption{Map of the candidate cluster pair NGC1836-BRHT4b; the stellar system NGC1839 is also visible within the field of view. The centre of each cluster is indicated with an arrow.}
\label{map3}
\end{figure}

The data-set used in this paper consists of a combination of $V_{\rm HIGH}$ and $I_{\rm BESSEL}$ images obtained with the wide-field imager FORS2 at the 
Very Large Telescope (Prop ID: 090.D-0348, PI: Mucciarelli). Observations were obtained by using the $2k\times4k$ pixels MIT Red-optimized CCD mosaic in the high-resolution mode 
($\sim 0.12\arcsec$ pixel$^{-1}$), which yields a total field of view (FOV) of about $3.4\arcmin \times 3.4\arcmin$. 
In all cases, candidate cluster pairs were centered in chip$\#1$ (Figures~\ref{map1}-\ref{map3}). In the FORS2 images 
targeting SL385-SL387 and NGC1836-BRHT4b we also observe NGC1922 and NGC1839 respectively (see Figures.~\ref{map2},~\ref{map3}).
The projected distances of these two clusters from the candidate cluster pairs is larger than $120\arcsec$.
According to \citet{Bhatia1990} and \citet{SugimotoMakino1989}, binary clusters with such large separations may become detached by external tidal
forces in relatively short timescales.
As a consequence we will consider them as unbound from the nearby candidate pairs.
However, they will be included in the following analysis.

For SL349-SL353, 16 images have been obtained both for $V_{\rm HIGH}$ and $I_{\rm BESSEL}$ with a combination of 
10 long exposures ($t_{\rm exp}=240$s and $t_{\rm exp}=280$s for $V_{\rm HIGH}$ and $I_{\rm BESSEL}$ respectively) 
and six images for each band with $t_{\rm exp}=10$s. 17 images have been acquired in $I_{\rm BESSEL}$ band and 16 
in the $V_{\rm HIGH}$ for SL385-SL387: $10 \times 240$s + $7\times 10$s in $I_{\rm BESSEL}$ and $10 \times 280$s + $6\times 10$s in $V_{\rm HIGH}$. 
In the case of NGC1836-BRHT4b a total of 15 images have been acquired in the $V_{\rm HIGH}$ band, 10 with exposure time $t_{\rm exp}=100$s 
and five with $t_{\rm exp}=10$s, and a total of 15 images in the $I_{\rm BESSEL}$ with the same combination of short and long exposures. 
A dither pattern of $\sim25\arcsec$ has been adopted for all targets to allow for a better reconstruction of the point-spread function (PSF) and to avoid CCD blemishes and artifacts. 
Master bias and flat-fields have been reconstructed by using a large number $(>20)$ of calibration frames. 
Then scientific images have been corrected for bias and flat-field by using standard procedures and tasks contained in the Image Reduction and Analysis 
Facility (IRAF).\footnote{IRAF is written and supported by the National Optical Astronomy Observatories (NOAO), which is operated by the Association of Universities for Research in Astronomy, Inc., 
under a cooperative agreement with the National Science Foundation.}

Following the same approach as in \citet{Dalessandro2015a}, the photometric analysis has been performed independently for each image and chip by using 
\nobreak{DAOPHOT\,IV} \citep{Stetson1987}. 
For each frame we selected several tens of bright, not saturated, and relatively isolated stars to model the PSF. For each chip the best PSF model was then applied 
to all sources at $3\sigma$ above the background by using \textrm{DAOPHOT/ALLSTAR}. We then created a master list of stars composed by sources detected at least in four frames. 
In each single frame, at the corresponding positions of the stars present in the master list, a fit was forced with \textrm{DAOPHOT/ALLFRAME}
\citep{stetson1994} 
For each star, different magnitude estimates in each filter were homogenized and their weighted mean and standard deviation were 
finally adopted as star magnitudes and photometric errors. 
Instrumental magnitudes were transformed to the Johnson/Cousin standard photometric system by using the stars in common 
with the catalog of \citet{Zaritsky2004} as secondary photometric standards. A few hundreds stars were found in the FOV of each candidate pair spanning the entire color range. 
Instrumental coordinates (x,y) were reported to the absolute ($\alpha$, $\delta$) system by using the stars in common with 2MASS 
and the cross-correlation tool \textrm{CataXcorr}\footnote{CataXcorr is a code aimed at cross-correlating catalogs 
and finding solutions, developed by P. Montegriffo at INAF - Osservatorio Astronomico di Bologna.}.

\begin{figure*}
\includegraphics[width=140mm]{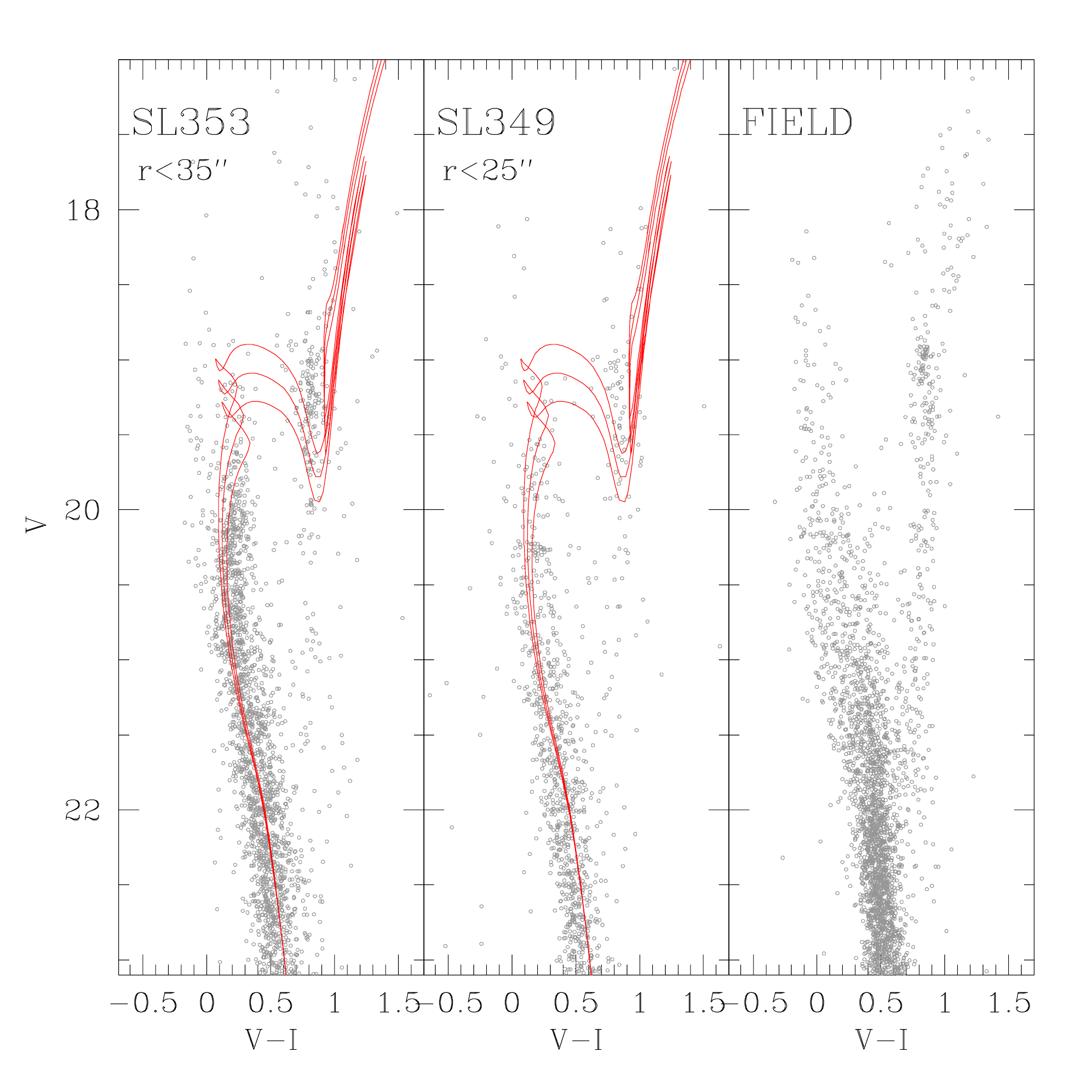}
\caption{From left to right: color-magnitude diagrams of the clusters SL353 and SL349 and of the field stars. The area sampled by the field
CMD is $\sim2.4$ arcmin$^2$, thus more than 2.5 times the area covered by the CMDs of the two clusters.
In the first two panels best-fit isochrones as well as minimun and maximum age models providing an acceptable fit are shown.}
\label{cmd_1}
\end{figure*}

\begin{figure*}
\includegraphics[width=140mm]{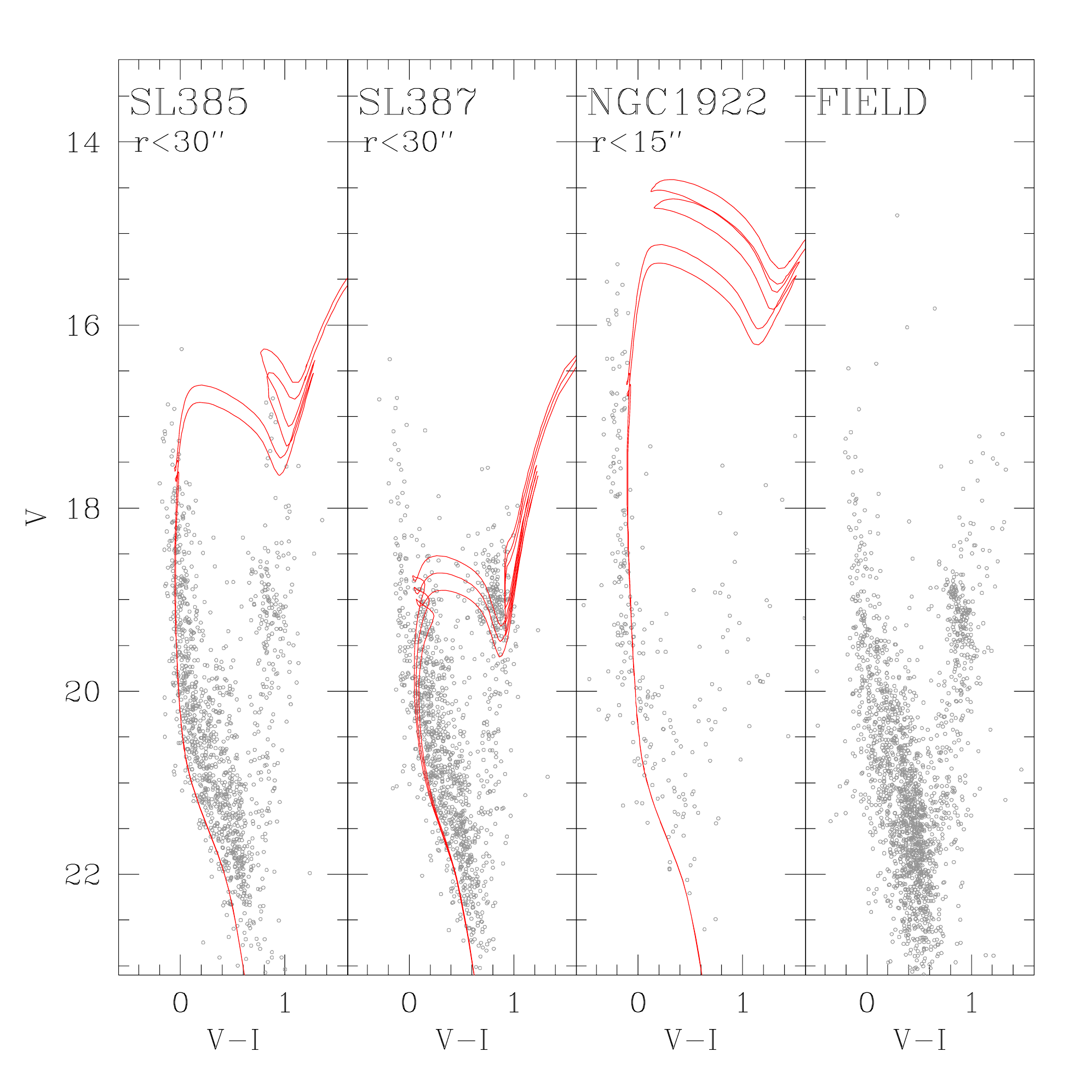}
\caption{From left to right: color-magnitude diagrams of the clusters SL385, SL387, NGC1922, and of the field stars. 
The area sampled by the field CMD is $\sim1.3$arcmin$^2$.
As in Figure\ref{cmd_1}, best-fit isochrones and minimun/maximum age models providing an acceptable fit are shown.}
\label{cmd_2}
\end{figure*}

At this stage, the catalogs obtained for each chip are on the same photometric and astrometric system. 
They have been combined to form a single catalog for each candidate pair. Stars in common between different pointing have been used to 
check for the presence of residuals in the calibration procedure. The resulting color-magnitude diagrams (CMDs) are shown in Figs~\ref{cmd_1}-\ref{cmd_3}.

\section{Age estimates}
In order to constrain the possible binarity of the candidate pairs in our sample, we will use in the following three main diagnostics (see
Introduction). 
In this Section, we start by deriving the cluster ages. 
Ages of stellar systems in pairs can provide important clues about their formation.
Clusters with similar ages likely formed from the same molecular cloud, while systems with significantly different ages are more likely
unbound or the result of a capture event.
 
The ages of the clusters were derived by comparing CMDs with a set of PARSEC isochrones  \citep{Bressan2012}. 
For all clusters we adopted a metallicity of $Z=0.006$ ([Fe/H]$\sim-0.4$), which is compatible with high-resolution spectroscopic estimates obtained 
for intermediate and young GCs in the LMC \citep[see for example][]{Mucciarelli2008,Mucciarelli2012}, 
a true distance modulus $(m-M)_0=18.55$ and reddening $E(B-V)=0.08$, which are compatible with typical values reported in the literature (see
for example \citealt{inno2016} and \citealt{haschke2011}).

In the following we list the results obtained for each candidate pair.

\begin{table}
\begin{center}
\caption[Ages.]{Estimated ages and gravity center of the clusters, see Sections 3 and 4.1 for details} 
\label{Ages}
\begin{tabular}{ccccc}
\hline\hline
Cluster  &   t                      & $\alpha$	& $\delta$ \\
         & (Myr)   		        & (h:m:s)	& ($\deg$:$\arcmin$:$\arcsec$) \\
\hline
SL353    & 1000 $\pm$ 120         & 05:17:07.938 & -68:52:24.51 \\
SL349    & 1000 $\pm$ 120 	        & 05:16:54.524 & -68:52:35.61 \\
\hline
SL385    & 240 $\pm$ 15 	        & 05:19:25.241 & -69:32:27.99 \\
SL387    & 740$^{+150}_{-120}$        & 05:19:33.686 & -69:32:32.62 \\ 
NGC1922 &  90 $\pm$ 10         & 05:19:50.353 & -69:30:01.04 \\
\hline
NGC1836 & 400 $\pm$ 50  	        & 05:05:35.700 & -68:37:42.56 \\ 
NGC1839 & 140 $\pm$ 15 	        & 05:06:02.596 & -68:37:43.13 \\
BRHT4b   & 140 $\pm$ 15 	        & 05:05:40.572 & -68:38:14.50 \\
\hline 
\end{tabular}
\end{center}
\end{table}

\smallskip

{\it SL349 - SL353} - To minimize the impact of field star contamination, which can significantly affect the age determination,  
we used only stars located at distance $r<35\arcsec$ and $r<25\arcsec$ from the gravity centers of 
SL349 and SL353 respectively (see Section~4.1). 
The resulting CMDs are shown in Fig.~\ref{cmd_1} (left and middle panels). For comparison we also show the CMD of stars 
located at a distance $r>200\arcsec$ from both clusters, 
which are representative of the surrounding field population. We find that the CMDs of SL349 and SL353 are best reproduced 
by models with ages $t=1.00\pm0.12$ Gyr. 
Isochrones nicely reproduce the main sequence shape as well as the MSTO and blue-loop star luminosity, which represents 
a stringent constraint to the overall fit. 
These ages are larger than what obtained by \citet{Dieball2000} who find $t=550\pm110$ Myr for both systems 
based on very shallow CMDs that do not reach the cluster MSTOs, and by \citet{Piatti2015} who estimates $t=500$ for SL349
by using VISTA near-IR observations.
On the contrary, they are broadly compatible with results obtained by \citet{Bica1996}, 
who classified both clusters in {\it category V}, i.e. in the age range 800-2000 Myr, by using UBV integrated colors.

\smallskip

{\it SL385 - SL387 - NGC1922} - The age estimates for these clusters were performed by using stars at distances from 
the cluster centers $r<30\arcsec$ for SL385 and SL387 and $r<15\arcsec$ for NGC1922 (see Fig.~\ref{cmd_2}). 
Figure~\ref{cmd_2} also shows the CMD of stars 
located at a distance $r>150\arcsec$ from all clusters for comparison. 
The CMD of SL385 shows a group of bright stars at $V\sim17$ mag and $(V-I)\sim1$ mag  mainly 
distributed in the innermost $10\arcsec$ and thus likely cluster members. 
In addition we note that this group of bright stars is not
present in either the CMD of the neighbor clusters nor in that of the surrounding field.
The position of these stars and the extension of the main sequence can be nicely fit by models with 
$t=240\pm15$ Myr. This estimate is compatible with that obtained by \citet{Piatti2015}. 
SL387 appears to be older than SL385. We find a best-fit age $t=740^{+150}_{-120}$ Myr. 
These results are in good agreement within the errors with estimates obtained by \citet{Vallenari1998} for both systems 
based on resolved CMDs. On the contrary, \citet{Bica1996}  
classify both GCs in {\it category IVA}, which includes clusters with age t$=200-400$ Myr. 
The CMD of NGC1922 shows an extremely pronounced bright extension of the main sequence which suggests a very young age for this system. 
We find it can be well reproduced with models representing an age of about $t=90\pm 10$ Myr.

\begin{figure*}
\includegraphics[width=140mm]{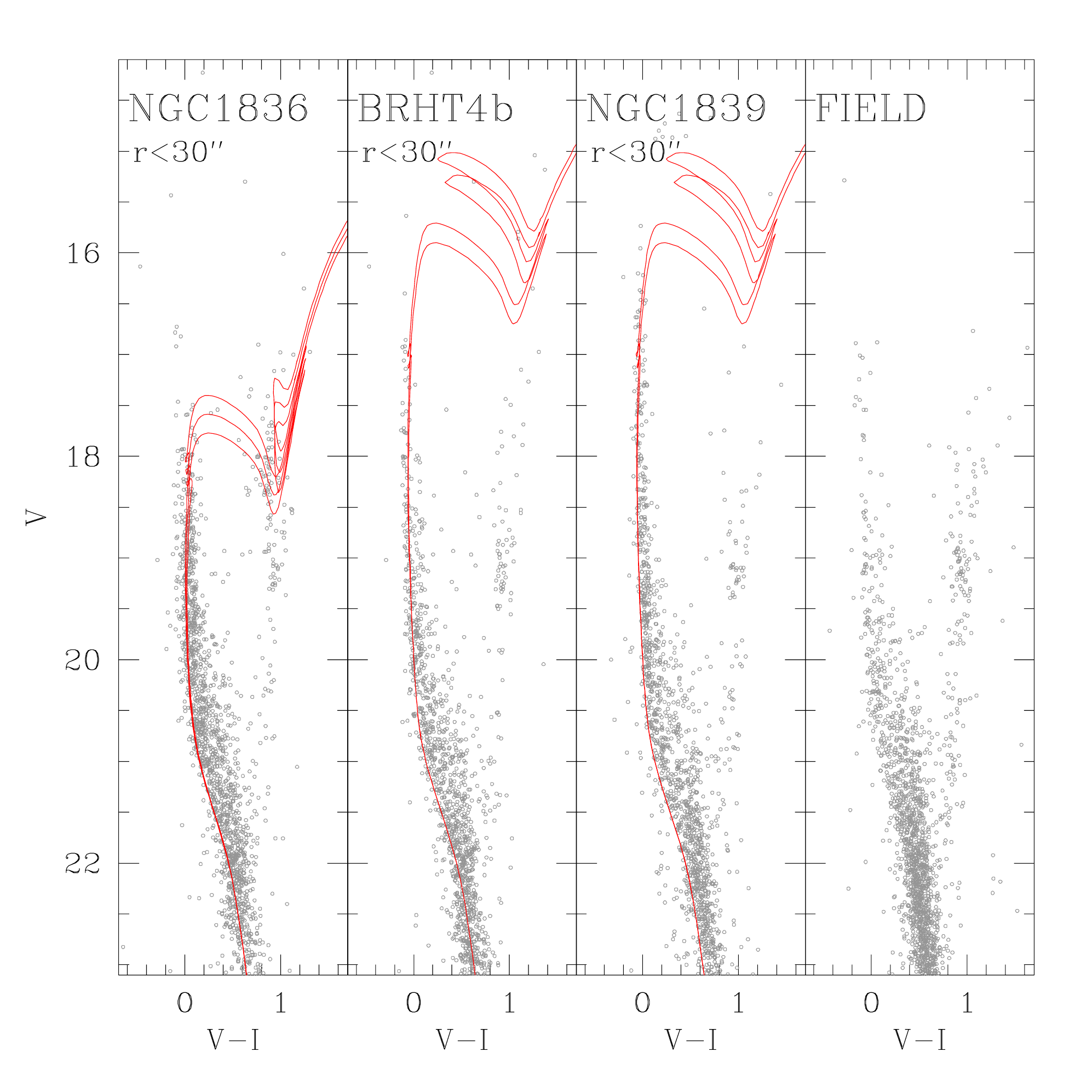}
\caption{As in Figure~\ref{cmd_2}, but for NGC1836, BRHT4b, NGC1839 and field stars. 
The area sampled by the field CMD is $\sim2.1$arcmin$^2$.}
\label{cmd_3}
\end{figure*}

\smallskip

{\it NGC1836 - BRHT4b - NGC1839} - For these three systems theoretical models were compared to stars located at distance 
$r<30\arcsec$ from their gravity centers (Fig.~\ref{cmd_3}). We also show a field CMD with stars 
located at a distance $r>150\arcsec$ from the three clusters. 
At a first inspection of the CMDs, NGC1836 appears older than the other two clusters, which in fact show a MS extending up 
to $V\sim16$. Also, the CMD of NGC1836 shows a clump of
data points at $V\sim18$ mag and $(V-I)\sim1$ mag that is
not evident in the CMDs of BRHT4b and NGC1839, a further indication of
its older age. Indeed we find that NGC1836 can be fitted with an isochrone of age $t=400\pm50$ Myr, 
while both BRHT4b and NGC1839 are compatible with $t=140\pm15$ Myr. 
These values are larger than what was found by \citet{Bica1996} who obtained ages in the range 70-200 Myr and 30-70 Myr for NGC1836 and 
NGC1839 respectively. No estimates are available for BRHT4b.

We have verified that a variation of the adopted metallicity $\Delta$[Fe/H]$\sim\pm0.4$ dex have an impact on the derived 
ages of $\sim 10\%$. 
Ages obtained for all clusters are summarized in Table~\ref{Ages}.
Note that errors give the minimum and maximum age providing an acceptable fit to the CMD.

\section{Density profiles and cluster parameters}
The other two diagnostics used in the present analysis to asses the binarity of the candidate pairs 
are mainly related to the cluster structural and morphological 
properties. 
In the following two Sections we will derive density profiles, the structural parameters of the clusters 
and constrain the effect of the tidal environment on their properties. 

\subsection{Centers of gravity and projected density profiles}
\label{Sect_densprof}

As a first step to compute the cluster density profiles, 
we derived the center of gravity, $C_{\rm grav}$, for each system by 
averaging the positions $\alpha$ and $\delta$ of properly selected stars and using an
iterative procedure \citep[see for example][]{Lanzoni2007,Dalessandro2013}. 
Only stars with $V \leq 21$ were used to avoid spurious effects due to incompleteness. 
For each target  we derived centers of gravity for different radial selections typically ranging 
from $\sim10\arcsec$ to $\sim20\arcsec$ (the only exception is SL439 for which 
an estimate of $C_{\rm grav}$  was obtained also using stars at a distance of $35\arcsec$) 
depending on the apparent 
extension of the systems and on the relative proximity to nearby clusters. 
We obtained a minimum of three to a maximum of five different estimates of the center for each cluster. 
$C_{\rm grav}$ was then obtained as the average of these values 
and the error as the standard deviation, which results to be typically of $\sim1\arcsec$.  
The centers thus derived for each cluster are listed in Table~\ref{Ages}.

The projected number density profiles were then determined by using direct star counts.  
Using the procedure described in \citet{Dalessandro2013}, we divided the selected regions into several concentric annuli of variable width 
(the exact number differs from cluster to cluster depending on their extent) centered on $C_{\rm grav}$ and suitably split 
in an adequate number of sub-sectors (in the range 2-4) depending on the portion of the field of view actually sampled. 
In order to minimize the contamination from nearby clusters, for each stellar system in the proposed pairs we considered 
only sub-sectors located in the opposite direction to the nearest GC (see Figs~\ref{map1}, \ref{map2} and \ref{map3}).
Number counts were calculated in each sub-sector and the corresponding densities 
were obtained dividing them by the sampled area. The number density of each annulus was then defined as the average of 
the sub-sectors densities, and its standard deviation was computed from the variance among the sub-sectors.

\begin{figure*}
\includegraphics[width=\textwidth]{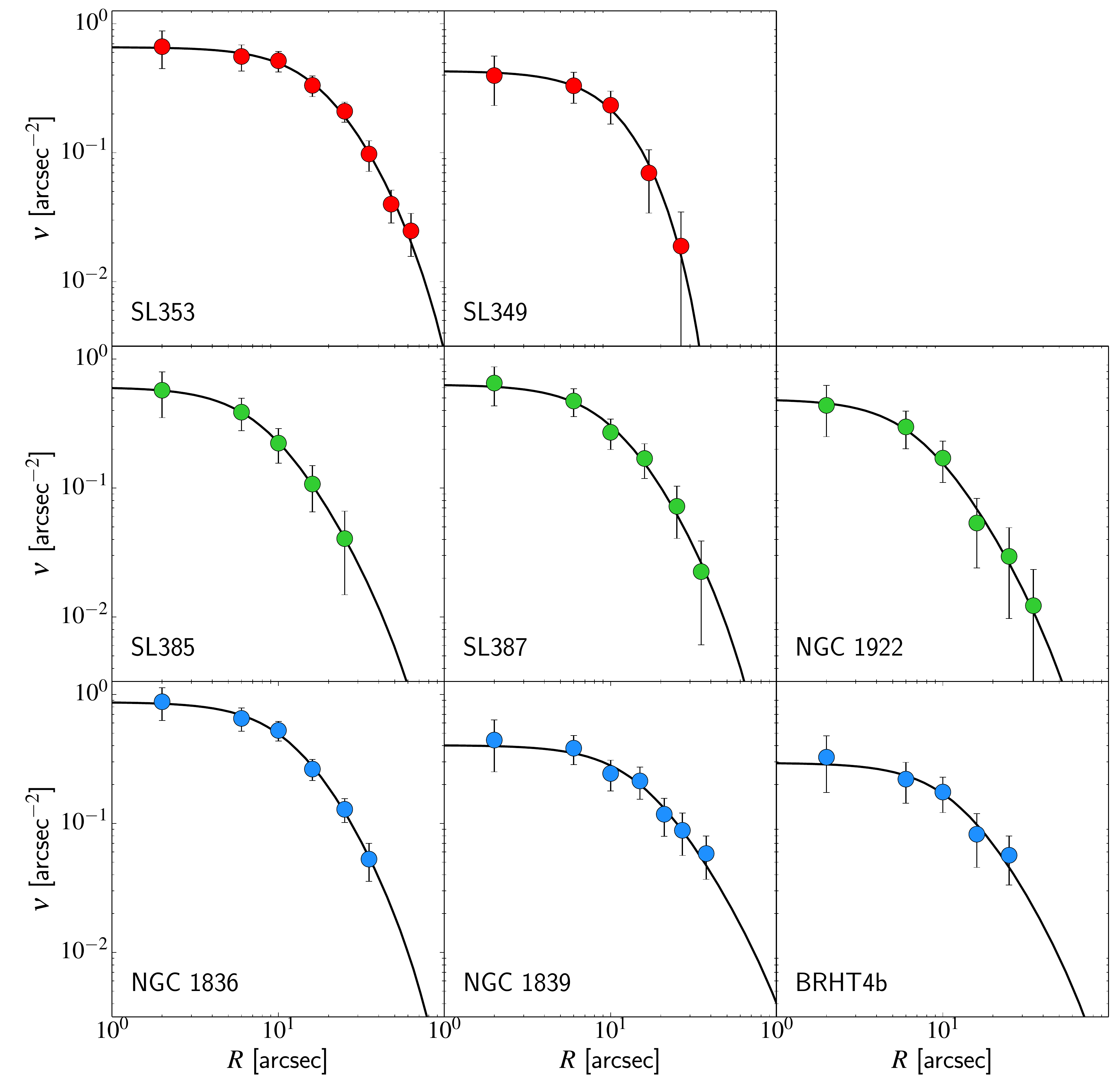}
\caption{Number density profiles of the clusters in our sample, as indicated by the labels in the panels. Solid lines correspond to the best-fit King (1996) model 
fits, and the data are indicated with circles. Each row corresponds to a different group of clusters, 
identified also by the different colors of the points in the panels. For each data-point, vertical error bars are shown. 
The vertical and horizontal axes span the same range in every panel, to allow a more direct comparison among the clusters.}
\label{NDprofiles}
\end{figure*}

Finally, for each system the background density contribution was estimated by using the density measurements of the outermost 
annuli. 
We notice that the background densities obtained in the outskirts of the clusters in the same field of view are consistent 
with each other. We then subtracted these values from the corresponding observed density profiles (see Fig.~\ref{NDprofiles}). 

\begin{table}
\begin{center}
\caption[Best-fit parameters.]{Best-fit parameters. For each cluster, listed in the first column, we provide the values of the best-fit parameters obtained by fitting King models to the number density profiles (see Section~\ref{Sect_fits}): the concentration parameter $W_0$, the King scale radius $r_0$  (expressed in arcsec), and the central value of the number density, $\nu_0$ (expressed in number per arcsec$^2$). We also indicate the formal errors on the parameters \citep[see][]{ZBV2012}.} 
\label{Tab_Best_Fit_Parameters}
\begin{tabular}{ccccccccccc}
\hline\hline
Cluster & $W_0$  & $r_0$    & $\nu_0$ \\
        &        & (arcsec) & (arcsec$^{-2}$) \\
\hline
SL353    & $4.78^{+1.48}_{-2.26}$ & $19.36^{+6.55}_{-5.69}$  & $0.658\pm0.003$ \\
SL349    & $1.88^{+3.02}_{-1.73}$ & $16.63^{+2.72}_{-11.37}$ & $0.430\pm0.005$ \\
\hline
SL385    & $5.92^{+3.94}_{-5.17}$ &  $8.56^{+4.60}_{-3.67}$  & $0.605\pm0.009$ \\
SL387    & $5.37^{+2.38}_{-4.65}$ & $10.73^{+4.62}_{-5.01}$  & $0.634\pm0.007$ \\ 
NGC1922 & $6.11^{+3.02}_{-5.10}$ &  $7.57^{+4.47}_{-3.46}$  & $0.489\pm0.008$ \\
\hline
NGC1836 & $5.28^{+2.54}_{-3.72}$ & $12.69^{+3.71}_{-4.80}$  & $0.871\pm0.006$ \\ 
NGC1839 & $6.20_{-0.48}^{+0.42}$ & $16.01_{-3.63}^{+4.97}$  & $0.404\pm0.002$ \\
BRHT4b   & $5.90_{-0.94}^{+0.70}$ & $12.62_{-4.34}^{+7.78}$  & $0.295\pm0.003$ \\
\hline 
\end{tabular}
\end{center}
\end{table}

\subsection{Best-fit dynamical models}
\label{Sect_fits}

We analyzed the number density profiles of the clusters in our sample by means of dynamical models.

We considered spherical, isotropic \citet{King1966} models\footnote{To compute these models we used the 
code \limepy\ introduced by \citet{GielesZocchi2015}, by fixing the value of the truncation parameter $g = 1$. 
See \citet{King1966} and \citet{GielesZocchi2015} for more details on the models and their calculation.} 
and we fit them to the density profiles calculated as described in Section~\ref{Sect_densprof}. 
We determine two fitting parameters: the structural parameter $W_0$,
(this parameter is often referred to as concentration), and a scale radius, $r_0$, sometimes called King radius. 
A third parameter, depending on these two, is the central number density $\nu_0$, which is needed to vertically scale 
the model profiles to match the observations; this parameter is related to the total number of stars belonging to the cluster. 

The best-fit parameters are determined by minimizing the quantity:
\begin{equation}
\chi^2 = \sum_{i = 1}^{N} \left[ \frac{N_i - \nu_0 \, \hat{\nu}(R_i/r_0)}{\delta N_i} \right]^2 \ ,
\label{chisq}
\end{equation}
where $R_i$, $N_i$, and $\delta N_i$ are the radial position, number density and number density error for each of the $N$ points 
in the number density profile of each cluster. The quantity $\hat{\nu}(R_i/r_0)$ is the projected number density of the model, normalized to 
its central value. The central number density $\nu_0$ is obtained as

\begin{equation}
\nu_0 = \left[\sum_{i = 1}^{N} \frac{N_i \, \hat{\nu}(R_i/r_0)}{\delta N_i^2} \right]
\left[ \sum_{i = 1}^{N} \frac{\hat{\nu}(R_i/r_0)}{\delta N_i^2} \right]^{-1} \ ,
\end{equation}

for each pair of values $(W_0, r_0)$. 

The best-fit parameters obtained from this fitting procedure are given in Table~\ref{Tab_Best_Fit_Parameters}, and the best-fit profiles are shown 
in Fig.~\ref{NDprofiles}. The models appear to reproduce the observed profiles well, over their radial extent.

The profile slope in the outermost radial bins for clusters NGC1839 and BRHT4b is quite shallow. For these profiles, the models providing a 
good fit turn out to 
be extremely concentrated and tend to have as best-fitting values for $W_0$ and for $r_0$ the largest 
values that we explored to calculate the $\chi^2$ function 
defined in equation~(\ref{chisq}) and $r_t\sim3000\arcsec$.  
Thus, in order to determine the best fit, we imposed the truncation radius $r_t$  to be equal to $r_J$ (see Section~\ref{Sect_Tides}).
This choice is a compromise between having an appropriate radial 
range for the density profiles of the clusters and still providing an adequate description of the data points by the models. 
The best-fit models of these clusters, therefore, need to be considered with caution.

\section{Characterization of the cluster density maps}
\subsection{Effects of the tidal environment}
\label{Sect_Tides}

To understand whether these clusters are gravitationally bound and possibly tidally interacting, it is necessary to 
determine their Roche filling conditions. 
To do so, we estimated the Jacobi radius $r_{\rm J}$ of each cluster in our sample, and we compared it with the truncation
radius $r_{\rm t}$
obtained by the best-fit King models. For each cluster, the Jacobi radius may be calculated in an approximate way as:

\begin{equation}
r_{\rm J} = \left( \frac{G M}{\xi \, \Omega^2} \right)^{1/3} \ ,
\end{equation}

\noindent where $M$ is the mass of the cluster, $G$ is the gravitational constant, $\Omega$ is the orbital frequency of the cluster in the LMC, 
and $\xi = 4 -\kappa^2/\Omega^2$, with $\kappa$ the epicyclic frequency (for details see \citealt{BertinVarri2008}). 
For simplicity, we describe the potential of the LMC by means of a spherical Plummer model \citep[as done, for example, by][]{BekkiChiba2005} 
with scale length $b = 2.6$~kpc; such an assumption allows us to specify $\xi$ as a simple function of the galactocentric distance $R_0$:

\begin{equation}
\xi(R_0)= \frac{3\,R_0^2}{b^2+R_0^2} \, . 
\label{nu}
\end{equation}

To estimate the total mass of a cluster we considered the corresponding best-fit King model, and we calculated its total luminosity 
in $V$-band 
by opportunely scaling it to match the central surface brightness (measured directly on the images), $\mu_{\rm V,0}$. 
We then converted this to a mass estimate by multiplying it by the $V$-band mass-to-light ratio $M/L$ appropriate for the age of the clusters and their
metallicity \citep{Maraston1998}\footnote{$M/L$ values are listed at the following link http://www-astro.physics.ox.ac.uk/$\sim$maraston/SSPn/ml/ml\textunderscore SSP.tab}.

To calculate the value of $\Omega$, we rely on the measures of the LMC centre and rotation curve recently obtained 
by \citet{vdMSahlmann2016} as a result of their analysis of \textit{HST} and \textit{Gaia} proper motions (see fourth column of their Table~2). 
They describe the rotation curve of the LMC as a function of the distance from the centre ($\alpha_0,\delta_0 = 79.37,-69.58$): 
the circular velocity increases linearly up to a velocity of 78.9 km~s$^{-1}$ at a distance of about 2.6 kpc from the 
centre\footnote{We note that we are using this value as the scale length $b$ to describe the potential of the LMC introduced above.}, 
and then remains flat outwards. The clusters in our sample are in the radial range where the circular velocity is linearly increasing with the distance, 
therefore for all of them $\Omega = 0.03$ km$\,$s$^{-1}$pc$^{-1}$. 

\begin{figure}
\includegraphics[width=\columnwidth]{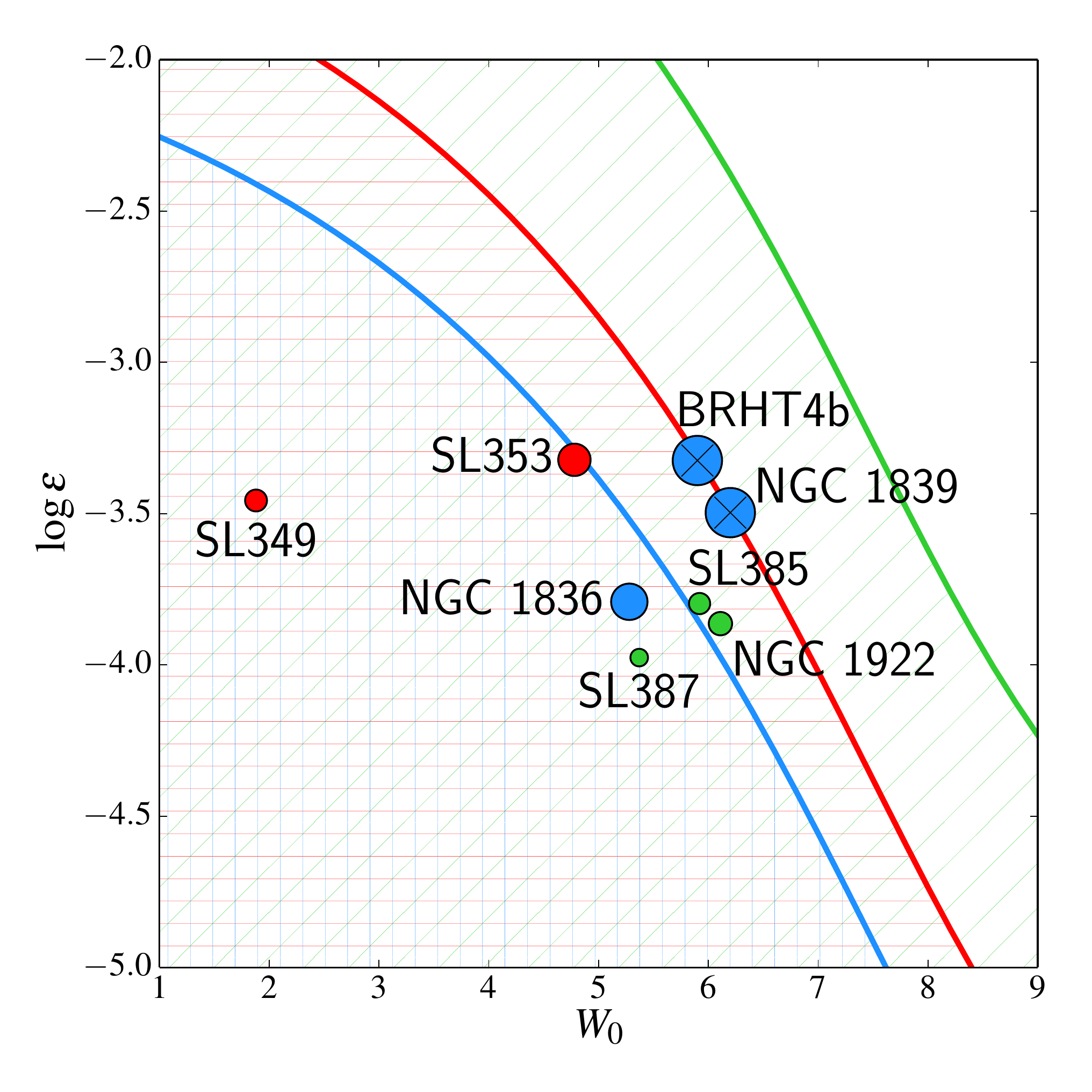}
\caption{Tidal strength parameter $\varepsilon$ as a function of $W_0$. 
The blue, red and green lines correspond to the critical values of the tidal strength parameter, as a function of $W_0$, for clusters located 
at the galactocentric distance corresponding to parameter values $\xi = 0.01,\, 0.16,\,0.55$ in a spherical host 
Plummer potential (for details, see main text). Any cluster located in the shaded region below a given line 
is underfilling its critical equipotential surface. 
The clusters in our sample are indicated with circles having the 
same colors used in Fig.~\ref{NDprofiles} and size representing the value of their filling factor $r_{\rm t}/r_{\rm J}$. Clusters NGC1839 
and BRHT4b are marked with a cross because their estimated truncation radius is chosen to be equal to their Jacobi radius.}
\label{logeW0}
\end{figure}

In Table~\ref{Tab_derived_quantities} we list the values of $r_J$ as well as the values of the quantities needed to 
obtain them and mentioned above. In the Table we also report the values of $r_{\rm t}/r_{\rm J}$ and 
$r_{\rm h}/r_{\rm J}$. These two quantities indicate the degree of filling of the Roche lobe of a given cluster. 
We notice that globular clusters having $r_{\rm h}/r_{\rm J}>0.10$ are usually considered to be tidally filling 
\citep[e.g., see][]{GielesBaumgardt2008,Baumgardt2010}. Based on the quantities derived in Table~\ref{Tab_derived_quantities} 
only the clusters in the NGC1836-BRHT4b-NGC1839 multiplet would be classified as tidally filling.

\begin{table*}
\begin{center}
\caption[Clusters properties.]{Relevant clusters properties. For each cluster, listed in the first column, 
we provide the values of several quantities: column (2) the galactocentric distance in kpc (assumed to be equal to the projected distance),
(3) the truncation radius determined from the best-fit King model, 
expressed both in arcsec and in pc, (4) the central surface brightness in V-band, (5) the V-band mass-to-light ratio 
in solar units, (6) the total mass in units of $10^4$ M$_{\odot}$, (7) the central mass density in unit of M$_{\odot}\,$pc$^{-3}$, 
(8) the Jacobi radius, both in arcsec and in pc, 
(9) the logarithm of the tidal strength parameter $\varepsilon$, (10) the ratio of the truncation radius to the Jacobi radius,
and (11) the ratio of the half-mass radius to the Jacobi radius. 
Quantities marked with an asterisk have been obtained by imposing, during the fitting procedure, 
the maximum possible value of the truncation radius (i.e., $r_t = r_j$).} 
\label{Tab_derived_quantities}
\begin{tabular}{ccc@{\hskip 0.05in}cccccc@{\hskip 0.05in}cccc}
\hline\hline
   (1)  &  (2)  & \multicolumn{2}{c}{(3)}         &    (4)          &  (5)  &  (6)&  (7)     & \multicolumn{2}{c}{(8)}         &     (9)            &     (10)              &         (11)            \\             
Cluster & $R_0$ & \multicolumn{2}{c}{$r_{\rm t}$} & $\mu_{\rm V,0}$ & $M/L$ & $M$ & $\rho_0$ & \multicolumn{2}{c}{$r_{\rm J}$} & $\log \varepsilon$ & $r_{\rm t}/r_{\rm J}$ & $r_{\rm h}/r_{\rm J}$ \\ 
        & kpc & arcsec & pc & mag$\,$arcsec$^{-2}$ & M$_{\odot}$/L$_{\odot}$ & $10^4$ M$_{\odot}$ & M$_{\odot}\,$pc$^{-3}$ & arcsec & pc  & &  \\
\hline 
SL353    & 0.62 & 186.96 &  45.41 & 19.7 & 0.62 & 4.04 & 35.63 &  434.97 & 105.65 & -3.32 & 0.43 & 0.08 \\
SL349    & 0.62 &  50.57 &  12.28 & 19.9 & 0.62 & 0.82 & 48.56 &  256.12 &  62.21 & -3.46 & 0.20 & 0.06 \\
\hline 
SL385    & 0.15 & 147.16 &  35.74 & 18.4 & 0.26 & 1.57 & 106.38 &  795.17 & 193.14 & -3.80 & 0.19 & 0.03 \\
SL387 & 0.16 & 137.81 & 33.47 & 18.6 & 0.58 & 3.86 & 160.50 & 1027.33 & 249.53 & -3.98 & 0.13 & 0.02 \\
NGC1922 & 0.19 & 145.04 &  35.23 & 17.9 & 0.17 & 1.35 & 123.96 &  645.27 & 156.73 & -3.86 & 0.22 & 0.03 \\
\hline 
NGC1836 & 1.25 & 155.13 &  37.68 & 18.4 & 0.37 & 3.99 & 104.94 &  285.68 &  69.39 & -3.79 & 0.54 & 0.10 \\
NGC1839 & 1.22 & 323.75$^*$ & 78.64$^*$ & 18.0 & 0.17 & 5.70 & 53.26 & 323.75 & 78.64 & -3.50 & 1.00$^*$ & 0.14 \\
BRHT4b   & 1.24 & 214.50$^*$ & 52.10$^*$ & 18.7 & 0.17 & 1.68 & 35.82 & 214.50 & 52.10 & -3.32 & 1.00$^*$ & 0.15 \\
\hline 
\end{tabular}
\end{center}
\end{table*}

We further explored this aspect by considering a dimensionless quantity introduced by \citet{BertinVarri2008} as one of the parameters of a 
family of triaxial dynamical models of stellar systems shaped by the tidal field of their hosting galaxy. This tidal strength parameter, 
$\varepsilon$, is defined as the ratio of the square of the orbital frequency of the cluster in the galaxy to the square of the dynamical 
frequency associated with its central mass density $\rho_0$:

\begin{equation}
\varepsilon = \frac{\Omega^2}{4 \pi G \rho_0} \ ,
\end{equation}

\noindent where $\rho_0$ is here determined from the best-fit King models, and $\Omega$ is obtained as described above. 
We emphasize that the following analysis is aimed exclusively at the characterization of the tidal effects associated with 
the host galaxy, and does not account for the tidal perturbations determined by any possible gravitational interaction between 
the members of a cluster pair.

The two-dimensional parameter space defined by the central concentration ($W_0$) and the tidal strength parameter is illustrated in Fig.~\ref{logeW0}. 
In such a diagram, for a given choice of the galactic potential and galactocentric distance $R_0$, we can identify configurations corresponding to 
the critical values of the tidal strength parameter (marked with solid lines), as a function of the central concentration 
parameter \citep[see][]{BertinVarri2008}. For a given tidal environment and value of $W_0$, the boundary of a critical configuration is 
defined by the last closed equipotential surface (i.e., such that $r_{\rm t}/r_{\rm J} \approx 2/3$, for details see \citealt{VarriBertin2009}, Sect. 2). 
From the bottom to the top, the lines represent the cases with $\xi = 0.01, 0.16$, and $0.55$, which, for simplicity, 
corresponds to the numerical average of the individual values of the parameter $\xi$ (see equation~\ref{nu})
resulting from the values of the galactocentric radii of the multiplets members (see Table~\ref{Tab_derived_quantities}, Col. 2). 
The hatched regions indicate configurations that are tidally underfilling, and, moving from the critical lines towards the bottom 
left corner of the parameter space, that are progressively less affected by the tidal perturbation. The clusters in our sample are 
indicated in the figure with circles of variable sizes; their diameter represents the value of their filling factor $r_{\rm t}/r_{\rm J}$. Clusters 
NGC1839 and BRHT4b are marked with a cross because their estimated truncation radius is chosen to be equal to their 
Jacobi radius (i.e., they correspond to overcritical configurations). The circles and their corresponding critical lines have the 
same colors used in Fig.~\ref{NDprofiles}. A better estimate of the LMC potential would provide a more accurate estimate 
of the tidal radii and of the tidal strength of these clusters.

\begin{figure*}
\includegraphics[width=175mm]{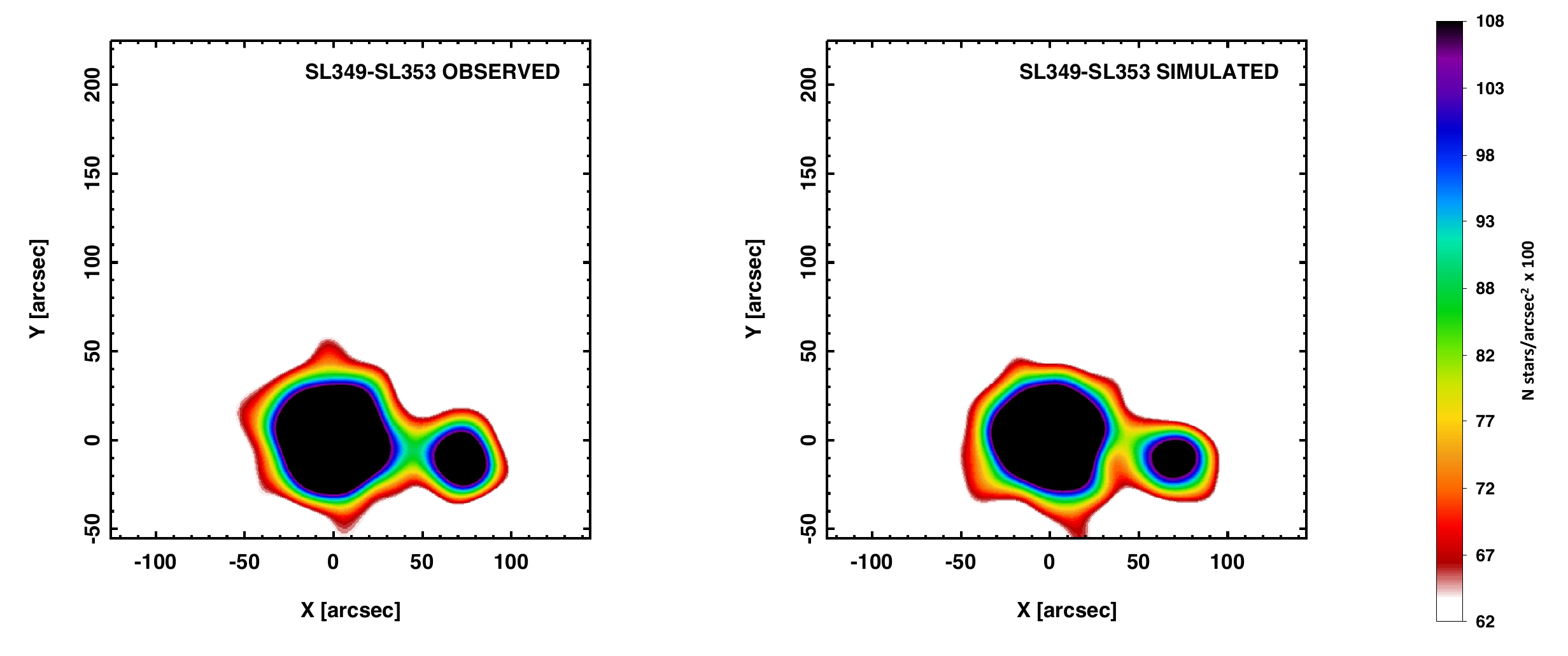}
\caption{Number density map for the clusters SL353 and SL349. The left panel shows the density map measured from the observations, 
the right panel the density map obtained from one of our 1000 simulated samples to reproduce this cluster pair. 
The color code presented in the bar on the right indicates the density scale.}
\label{comp}
\end{figure*}

Clusters SL349 and SL353 (indicated with red circles in the figure) have similar tidal strength but very different concentration; 
the first one, which is less massive and strongly underfilling, is well within the region indicating model tidal interactions, while SL353 
is relatively close to the condition of critical filling; 
in our sample, this pair is indeed characterized by the largest distance between members ($\sim$ 18 pc). 
Clusters SL387 and SL385 
(green circles in the figure) are located at a relative distance of $\approx 11$pc and they are both tidally underfilling 
(i.e., fall below the critical green line).
Finally, NGC1836 and BRHT4b (blue circles in the figure), which are at a relative distance of $\sim 10$ pc, have a very similar concentration
but different tidal strengths, with the first being underfilling and the second being overfilling.
It is important to clarify that BRHT4b and NGC1839 are overfilling by construction as we fixed $r_t = r_J$.
Also, it is worth noting that NGC1836 is almost critically filling its Roche lobe 
(as illustrated by its proximity to the critical blue line).

This analysis confirms that virtually all clusters in our sample are tidally underfilling. 
We also note that for all the clusters (except NGC1839 and BRHT4b, for which we do not have a reliable estimate for the truncation 
radius and we fixed it to be $r_{\rm t} = r_{\rm J}$), the Jacobi radius results to be larger than the truncation radius. 
For all candidate binary clusters, both the truncation radius and the Jacobi radius are estimated to be larger than the distance 
between them with the only exception being SL349, which has a very small truncation radius ($r_{\rm t} = 12.28$~pc).

In such a configuration, and assuming that the projected distances are compatible with the real ones,
the presence of intra-clusters stellar streams or bridges would be a strong indication that they  
are gravitationally bound as each one would fall within the Roche lobe 
(and in some cases even within the spatial truncation) of the companion. 
Very interesting are also the cases of SL353 and NGC1836, which are approaching conditions of critical filling, and therefore they 
are likely starting to loose stars trough their Roche lobes.

\subsection{Characterization of the intra-cluster over-densities}

To probe the spatial distribution of cluster pairs and the possible presence of interaction signatures, 
we analyzed their 2D density distributions.
The density analysis has been performed in the entire FORS2 FOV for each multiplet using only stars with $V\leq21$ in order to 
limit the impact of the background. 
The distribution of star positions was transformed into a smoothed surface density function through the use of a
kernel whose width has been fixed at $10\arcsec$ (see \citealt{Dalessandro2015b}). 
This procedure yields the surface density distribution shown as an example 
in Figure~\ref{comp} (left panel) for SL349-SL353. Each cluster appears to be quite spherical and a 
significant overdensity (a sort of bridge) between the two is clearly observed. 
This result is qualitatively compatible to what was found by \citet{Dieball2000}, however we do not confirm the elongation the authors
observed in SL353. We argue that such a discrepancy is likely due 
to the use of shallow photometry by \citep{Dieball2000}, which might be prone to low-number statistics and fluctuations in the distribution of bright stars.
 
In general, we find that all cluster pairs in our sample show evidence of intra-cluster overdensities.
These features could be an indication of an ongoing interaction between the clusters and therefore of their binarity. 
On the other hand however, they could be also due to projection effects. 

In order to constrain the nature of the observed intra-cluster overdensities, we used the best-fit King models described above.
For each pair of clusters, we generated 1000 simulated observations by sampling the distribution function of its best-fitting King model 
to randomly generate a set of stars. We locate the clusters at their relative positions, and we also simulated a uniform background, 
by using the background density we measured and subtracted from the number density profiles (see Section~\ref{Sect_densprof}). 
For each cluster, we used the value of the parameter $\nu_0$ obtained from the fitting procedure to scale the best-fit model density 
profiles: this allows us to obtain an estimate of the total number of stars observed in each cluster (by integrating the number density 
over the area). We then use this number to set the amount of stars to simulate in order to reproduce each cluster. 
After simulating each candidate cluster pair, we checked that the total number of stars in the clusters and the background 
are consistent with the total number of observed stars. 
Figure~\ref{comp} provides a comparison between the measured number density map in the field of view of SL349 - SL353 and the same map 
obtained by considering one of the simulated observations. The two panels show qualitatively similar density maps and in particular we observe that 
intra-cluster overdensities are clearly detectable also in our simulations.

\begin{figure}
\includegraphics[width=\columnwidth]{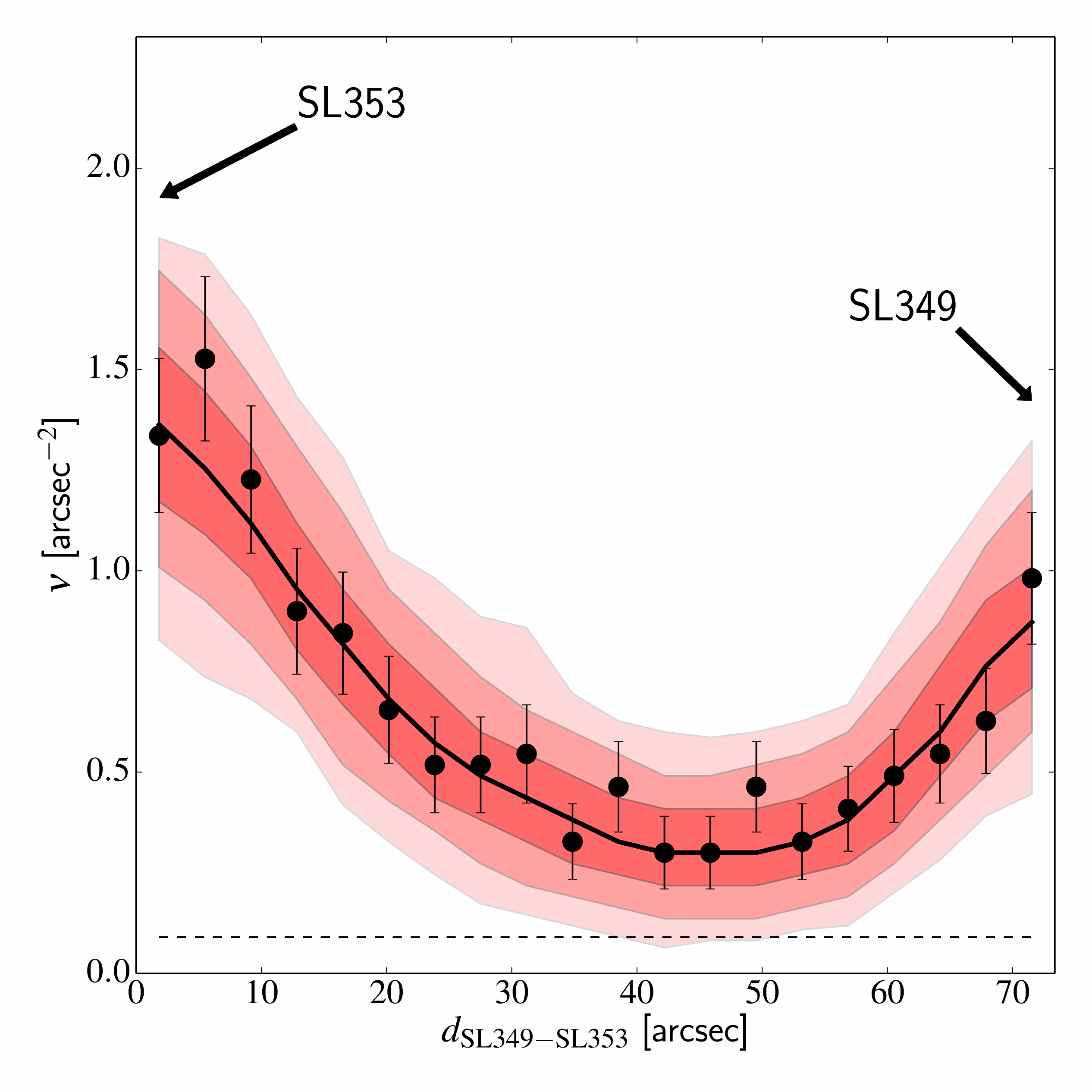}
\caption{Black points with vertical error bars 
 indicate the number density measured from the data along the line connecting clusters SL353 and SL349 on the plane of the sky. 
 The thick solid line represents the median of the distribution of the number
 density calculated from the 1000 random realization of the simulated observations in the same sky region, 
 the shaded areas correspond to 1$\sigma$, 2$\sigma$ 
 and 3$\sigma$ from the median. The dashed line indicates the uniform background density.}
\label{overdens_1}
\end{figure}

\begin{figure}
\includegraphics[width=\columnwidth]{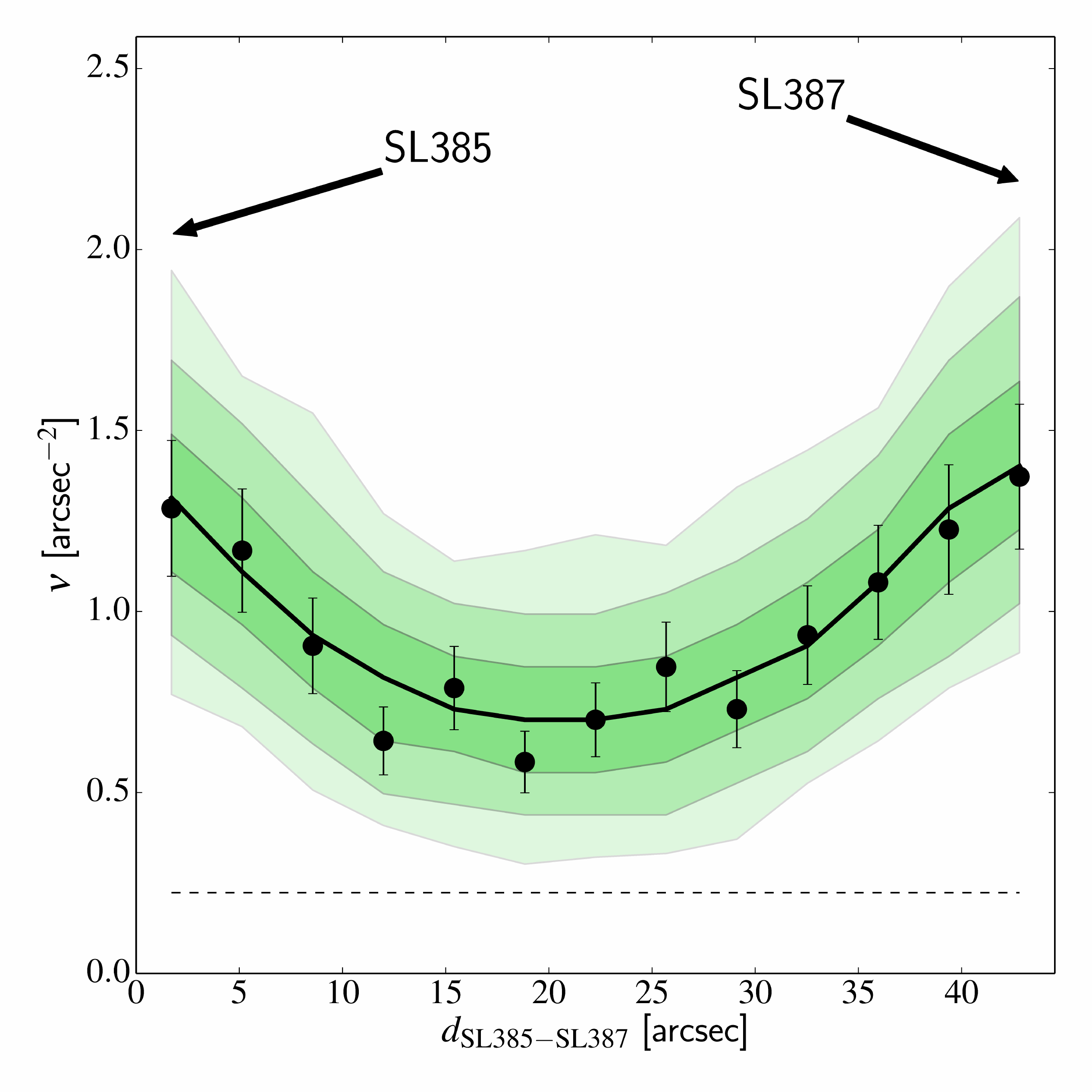}
\caption{Same as Fig.~\ref{overdens_1}, but for SL385 and SL387.}
\label{overdens_2}
\end{figure}

\begin{figure}
\includegraphics[width=\columnwidth]{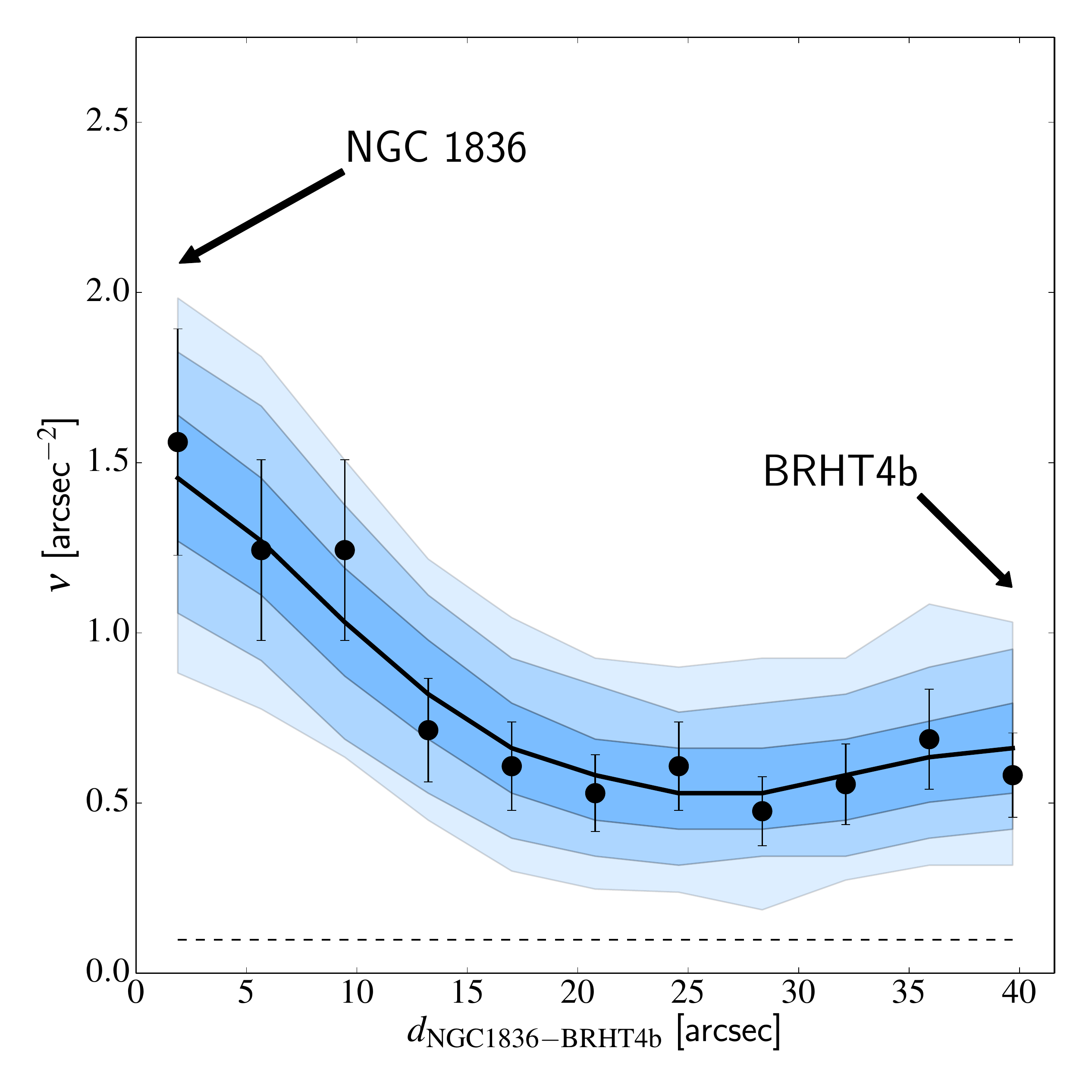}
\caption{Number density along the line connecting clusters NGC1836 and BRHT4b on the plane of the sky. The format of this figure is the same as Fig.~\ref{overdens_1}.}
\label{overdens_3}
\end{figure}

To provide a more quantitative comparison, for each simulated configuration we selected a rectangular region connecting the two candidate binary clusters,
with length equal to the distance between their centers and width of 10 arcsec. 
We divided this region into equal size bins and we calculated the number density of stars in each bin. 
Then, we computed the quantiles of the distribution in each bin, to obtain the median and the 1$\sigma$, 2$\sigma$ and 3$\sigma$ values 
for each bin. Figures~\ref{overdens_1}, \ref{overdens_2}, and \ref{overdens_3} show the result 
for the three candidate binary 
clusters. The thick solid line represents the median of the distribution of the number density, the shaded areas correspond to 1$\sigma$, 2$\sigma$ 
and 3$\sigma$ from the median. The dashed line indicates the uniform density that we assumed for the background. 
The black points, with their vertical error bars, represent the number density measured in the
corresponding bins in the observed distribution. 

The comparison between the observations and the simulations would suggest that the overdensity between clusters 
in the candidate pairs is consistent with them being close to each other in projection.
We do not observe any additional feature indicating ongoing strong tidal interactions between the members of the pairs.

\section{Discussion and Conclusions}

We presented a detailed photometric analysis of three candidate cluster pairs in the LMC with the aim of characterizing their properties and constraining their possible binarity. 
Specifically, we have derived their ages, determined their structural and morphological properties and investigated the possible presence of signatures of gravitational 
interactions between the members of a given pair.

We found that the members of the pair SL349 - SL353 share the same age ($t=1.00\pm0.12$ Gyr), 
thus suggesting that these systems possibly formed from the fragmentation of the same molecular cloud. 
Also BRHT4b and NGC1839 have similar ages ($t=140\pm15$ Myr). Given their projected distance however, it is unlikely that these systems form a true pair, while of course we
cannot exclude they have a common origin. 
SL385, SL387 and NGC1922 show significant age differences and therefore they 
likely formed in different clouds.

By means of simple, single-mass, isotropic King (1966) models, we have derived 
an estimate of the structural parameters 
of all star clusters in our sample.
In addition, we have also provided an approximate estimate of the critical equipotential surface and Jacobi (tidal) radius of the star clusters, 
as a zeroth-order tool to evaluate the extension of their Roche lobe, as determined by the interaction with the tidal field of  
the LMC. 
We wish to emphasize that our study has a number of limitations. First of all, our simple dynamical analysis is based exclusively on the interpretation 
of photometric data, which do offer only a very partial, and often degenerate, view of the internal properties of star clusters 
(e.g., see \citealt{ZBV2012,Zocchi2016}). 
Second, the methodology for the calculation of both the truncation 
and tidal radius of the clusters in our study do not take into account the effects of possible gravitational interactions between 
the members of a given pair. 
%Such a limitation is significant, but may be overcome relatively easily, by constructing an appropriate 
%theoretical framework in which (as in the case of binary stars or planets) two star clusters are described self-consistently, 
%as an energetically bound pair (Varri et al., in preparation). 
Third, the calculation of the Jacobi radius of the clusters in 
our sample is also based on a relatively crude estimate of their orbital frequencies which, in turn, relies on a particularly 
simplified description of the potential of the LMC and on the heavy assumption that the three-dimensional galactocentric distance 
of the clusters correspond to the distance in projection. 

None the less, the simple structural information we have determined for the clusters in our sample have allowed us to evaluate their degree of filling, 
therefore, to have a first assessment of the strength of the perturbation associated with the external tidal field. 
All clusters appear to be underfilling their Roche lobe, with NGC1836 and SL353 being relatively close 
to the condition of critical filling 
(see Table~3 and Fig.~9). 
For all candidate binary clusters, both the truncation radius and the Jacobi radius are estimated to be larger than the distance 
between them. In such a configuration, and assuming that the projected distances are compatible with the real ones,
the presence of intra-clusters stellar streams or bridges would be a strong indication that they  
are gravitationally bound as each one would fall within the Roche lobe 
(and in some cases even within the spatial truncation) of the companion. 

Indeed all clusters show evidence of intra-cluster over-densities.
However, 
it appears to be impossible with basically only photometric information to distinguish the case in which there is a 
genuine presence of ongoing mass-transfer
from the case in which the members are simply sufficiently close to each other so that the tenuous mass distribution in their outer 
regions appears, in projection, to be a joined mass distribution. 

Among the candidate cluster pairs, the case of SL349 - SL353 is particularly interesting.
In fact, \citet{Dieball2000}
measured the radial velocities of a sample of individual stars in both systems. While the sample
of member stars is small (4 and 5 for SL349 and SL353 respectively), the authors 
concluded that  
the clusters have very similar mean velocities ($v_{SL349} \sim 274 \pm 10$ km s$^{-1}$ and $v_{SL353} \sim 279 \pm 4$ km s$^{-1}$) 
and they could share a common center of mass. 
   
All these elements, coupled with the fact that SL349 and SL353 share the same age ($t\sim 1$ Gyr) 
contribute to form a dynamical interpretation according to which the clusters SL349 and SL353 might 
actually be the members of an energetically bound pair. It is therefore imperative to pursue a more detailed 
kinematical analysis of such clusters, 
without which, at present, we do not dare making a conclusive statement about the true nature of this pair.     
Such a degeneracy in the interpretation 
may be broken exclusively by coupling the currently available photometric information with an appropriate 
kinematical characterization covering the entire radial extension of the clusters, 
and, ideally, of any star in the intra-cluster region. 

\section*{Acknowledgements}
We thank the anonymous referee for the carefully reading of the paper and his/her useful suggestions.
ALV acknowledges support from the EU Horizon 2020 program (Marie Sklodowska-Curie Fellowship, MSCA-IF-EF-RI-658088).

%%%%%%%%%%%%%%%%%%%%%%%%%%%%%%%%%%%%%%%%%%%%%%%%%%



%\bibliographystyle{mnras}
%\bibliography{biblio.bib}

\begin{thebibliography}{}

\bibitem[Amaro-Seoane et al.(2013)]{AmaroSeoane13} 
Amaro-Seoane, P., Konstantinidis, S., Brem, P., \& Catelan, M.\ 2013, MNRAS, 435, 809 

\bibitem[Arnold et al.(2017)]{arnold2017} Arnold, B., Goodwin, S.~P., Griffiths, D.~W., \& Parker, R.~J.\ 2017, MNRAS, 471, 2498 

\bibitem[Barnes \& Hut(1986)]{BarnesHut1986} Barnes, J., \& Hut, P.\ 1986, Nature, 324, 446 

\bibitem[Baumgardt et al.(2003)]{Baumgardt2003} Baumgardt, H., Makino, J., Hut, P., McMillan, S., \& Portegies Zwart, S.\ 2003, ApJl, 589, L25 

\bibitem[Baumgardt et al.(2010)]{Baumgardt2010} Baumgardt, H., Parmentier, G., Gieles, M., \& Vesperini, E.\ 2010, MNRAS, 401, 1832 

\bibitem[Bekki et al.(2004)]{Bekki2004} Bekki, K., Beasley, M.~A., Forbes, D.~A., \& Couch, W.~J.\ 2004, ApJ, 602, 730 

\bibitem[Bekki \& Chiba(2005)]{BekkiChiba2005} Bekki, K., \& Chiba, M.\ 2005, MNRAS, 356, 680

\bibitem[Bertin \& Varri(2008)]{BertinVarri2008} Bertin, G., \& Varri, A.~L.\ 2008, ApJ, 689, 1005-1019 

\bibitem[Bhatia \& Hatzidimitriou(1988)]{BhatiaHatz1988} Bhatia, R.~K., \& Hatzidimitriou, D.\ 1988, MNRAS, 230, 215 

\bibitem[Bhatia(1990)]{Bhatia1990} Bhatia, R.~K.\ 1990, PASJ, 42, 757 

\bibitem[Bhatia et al.(1991)]{Bhatia1991} Bhatia, R.~K., Read, M.~A., Hatzidimitriou, D., \& Tritton, S.\ 1991, A\&AS, 87, 335 

\bibitem[Bica et al.(1996)]{Bica1996} Bica, E., Claria, J.~J., Dottori, H., Santos, J.~F.~C., Jr., \& Piatti, A.~E.\ 1996, ApJS, 102, 57 

\bibitem[Bressan et al.(2012)]{Bressan2012} Bressan, A., Marigo, P., Girardi, L., et al.\ 2012, MNRAS, 427, 127 

\bibitem[Br{\"u}ns \& Kroupa(2011)]{Bruns2011} Br{\"u}ns, R.~C., \& Kroupa, P.\ 2011, ApJ, 729, 69

\bibitem[Catelan(1997)]{Catelan1997} Catelan, M.\ 1997, ApJl, 478, L99  

\bibitem[Dalessandro et al.(2015a)]{Dalessandro2015a} Dalessandro, E., Ferraro, F.~R., Massari, D., et al.\ 2015, ApJ, 810, 40 

\bibitem[Dalessandro et al.(2015b)]{Dalessandro2015b} Dalessandro, E., Miocchi, P., Carraro, G., J{\'{\i}}lkov{\'a}, L., \& Moitinho, A.\ 2015, MNRAS, 449, 1811 

\bibitem[Dalessandro et al.(2013)]{Dalessandro2013} Dalessandro, E., Ferraro, F.~R., Massari, D., et al.\ 2013, ApJ, 778, 135 

\bibitem[de La Fuente Marcos \& de La Fuente Marcos(2009)]{delafuente2009} de La Fuente Marcos, R., \& de La Fuente Marcos, C.\ 2009, A\&A, 500, L13 

\bibitem[de Oliveira et al.(2000)]{deOliveira2000} de Oliveira, M.~R., Bica, E., \& Dottori, H.\ 2000, MNRAS, 311, 589 

\bibitem[De Silva et al.(2015)]{desilva2015} De Silva, G.~M., Carraro, G., D'Orazi, V., et al.\ 2015, MNRAS, 453, 106 

\bibitem[Dieball et al.(2000)]{Dieball2000} Dieball, A., Grebel, E.~K., \& Theis, C.\ 2000, A\&A, 358, 144 

\bibitem[Dieball et al.(2002)]{Dieball2002} Dieball, A., M{\"u}ller, H., \& Grebel, E.~K.\ 2002, A\&A, 391, 547 

\bibitem[Efremov(1995)]{efremov1995} Efremov, Y.~N.\ 1995, AJ, 110, 2757 

\bibitem[Elmegreen \& Elmegreen(1983)]{ElmegreenElmegreen1983} Elmegreen, B.~G., \& Elmegreen, D.~M.\ 1983, MNRAS, 203, 31 

\bibitem[Gardiner \& Noguchi(1996)]{GardinerNoguchi1996} Gardiner, L.~T., \& Noguchi, M.\ 1996, Journal of Korean Astronomical Society Supplement, 29, S93 

\bibitem[Goodwin \& Whitworth(2004)]{goodwin2004} Goodwin, S.~P., \& Whitworth, A.~P.\ 2004, A\&A, 413, 929 

\bibitem[Fall et al.(2005)]{fall2005} Fall, S.~M., Chandar, R., \& Whitmore, B.~C.\ 2005, ApJL, 631, L133 

\bibitem[Gavagnin et al.(2016)]{Gavagnin2016} Gavagnin, E., Mapelli, M., \& Lake, G.\ 2016, MNRAS, 461, 1276 

\bibitem[Gieles \& Baumgardt(2008)]{GielesBaumgardt2008} Gieles, M., \& Baumgardt, H.\ 2008, MNRAS, 389, L28 

\bibitem[Gieles \& Zocchi(2015)]{GielesZocchi2015} Gieles, M., \& Zocchi, A.\ 2015, MNRAS, 454, 576

\bibitem[Girardi et al.(1995)]{Girardi1995} Girardi, L., Chiosi, C., Bertelli, G., \& Bressan, A.\ 1995, A\&A, 298, 87 
 
\bibitem[Haschke et al.(2011)]{haschke2011} Haschke, R., Grebel, E.~K., \& Duffau, S.\ 2011, AJ, 141, 158 

\bibitem[Hatzidimitriou \& Bhatia(1990)]{HatzBhatia1990} Hatzidimitriou, D., \& Bhatia, R.~K.\ 1990, A\&A, 230, 11 

\bibitem[Holland et al.(1995)]{Holland1995} Holland, S., Fahlman, G.~G., \& Richer, H.~B.\ 1995, AJ, 109, 2061 

\bibitem[Hong et al.(2017)]{Hong2017} Hong, J., de Grijs, R., Askar, A., et al.\ 2017, MNRAS, 472, 67

\bibitem[Inno et al.(2016)]{inno2016} Inno, L., Bono, G., Matsunaga, N., et al.\ 2016, ApJ, 832, 176 

\bibitem[Kallivayalil et al.(2013)]{Kallivayalil2013} Kallivayalil, N., van der Marel, R.~P., Besla, G., Anderson, J., \& Alcock, C.\ 2013, ApJ, 764, 161 

\bibitem[King(1966)]{King1966} King, I.~R.\ 1966, AJ, 71, 64 

\bibitem[Kumai et al.(1993)]{Kumai1993} Kumai, Y., Basu, B., \& Fujimoto, M.\ 1993, ApJ, 404, 144 

\bibitem[Lanzoni et al.(2007)]{Lanzoni2007} Lanzoni, B., Dalessandro, E., Ferraro, F.~R., et al.\ 2007, ApJl, 668, L139 

\bibitem[Larsen(2000)]{larsen2000} Larsen, S.~S.\ 2000, MNRAS, 319, 893 

\bibitem[Lee et al.(1999)]{Lee1999} Lee, Y.-W., Joo, J.-M., Sohn, Y.-J., et al.\ 1999, Nature, 402, 55 

\bibitem[Leon et al.(1999)]{Leon1999} Leon, S., Bergond, G., \& Vallenari, A.\ 1999, A\&A, 344, 450 

\bibitem[Makino et al.(1991)]{Makino1991} Makino, J., Akiyama, K., \& Sugimoto, D.\ 1991, Ap\&SS, 185, 63 

\bibitem[Maraston(1998)]{Maraston1998} Maraston, C.\ 1998, MNRAS, 300, 872 

\bibitem[Minniti et al.(2004)]{Minniti2004} Minniti, D., Rejkuba, M., Funes, J.~G., \& Kennicutt, R.~C., Jr.\ 2004, ApJ, 612, 215 

\bibitem[Mucciarelli et al.(2008)]{Mucciarelli2008} Mucciarelli, A., Carretta, E., Origlia, L., \& Ferraro, F.~R.\ 2008, AJ, 136, 375 

\bibitem[Mucciarelli et al.(2012)]{Mucciarelli2012} Mucciarelli, A., Origlia, L., Ferraro, F.~R., Bellazzini, M., \& Lanzoni, B.\ 2012, ApJl, 746, L19 

\bibitem[Okumura et al.(1991)]{Okumura1991} Okumura, S.~K., Ebisuzaki, T., \& Makino, J.\ 1991, PASJ, 43, 781 

\bibitem[de Oliveira et al.(1998)]{deOliveira1998} de Oliveira, M.~R., Dottori, H., \& Bica, E.\ 1998, MNRAS, 295, 921 

\bibitem[Piatti et al.(2015)]{Piatti2015} Piatti, A.~E., de Grijs, R., Ripepi, V., et al.\ 2015, MNRAS, 454, 839 

\bibitem[Portegies Zwart \& Rusli(2007)]{PortegiesZwartRusli2007} Portegies Zwart, S.~F., \& Rusli, S.~P.\ 2007, MNRAS, 374, 931 

\bibitem[Priyatikanto et al.(2016)]{Pry2016} Priyatikanto, R., Kouwenhoven, M.~B.~N., Arifyanto, M.~I., Wulandari, H.~R.~T., \& Siregar, S.\ 2016, MNRAS, 457, 1339 

\bibitem[Stetson(1987)]{Stetson1987} Stetson, P.~B.\ 1987, PASP, 99, 191 

\bibitem[Stetson(1994)]{stetson1994} Stetson, P.~B.\ 1994, PASP, 106, 250 

\bibitem[Subramaniam et al.(1995)]{Subramaniam1995} Subramaniam, A., Gorti, U., Sagar, R., \& Bhatt, H.~C.\ 1995, A\&A 302, 86 

\bibitem[Sugimoto \& Makino(1989)]{SugimotoMakino1989} Sugimoto, D., \& Makino, J.\ 1989, PASJ, 41, 1117

\bibitem[Surdin(1991)]{Surdin1991} Surdin, V.~G.\ 1991, Ap\&SS, 183, 129 

\bibitem[Vallenari et al.(1998)]{Vallenari1998} Vallenari, A., Bettoni, D., \& Chiosi, C.\ 1998, A\&A, 331, 506 
 
\bibitem[van den Bergh(1996)]{vandenBergh1996} van den Bergh, S.\ 1996, ApJl, 471, L31 

\bibitem[van der Marel \& Sahlmann(2016)]{vdMSahlmann2016} van der Marel, R.~P., \& Sahlmann, J.\ 2016, ApJl, 832, L23 

\bibitem[Varri \& Bertin(2009)]{VarriBertin2009} Varri, A.~L., \& Bertin, G.\ 2009, ApJ, 703, 1911 

\bibitem[Zaritsky et al.(2004)]{Zaritsky2004} Zaritsky, D., Harris, J., Thompson, I.~B., \& Grebel, E.~K.\ 2004, AJ, 128, 1606 

\bibitem[Zocchi et al.(2012)]{ZBV2012} Zocchi, A., Bertin, G., \& Varri, A.~L.\ 2012, A\&A, 539, A65 

\bibitem[Zocchi et al.(2016)]{Zocchi2016} Zocchi, A., Gieles, M., H{\'e}nault-Brunet, V., \& Varri, A.~L.\ 2016, MNRAS, 462, 696 

\end{thebibliography}


%%%%%%%%%%%%%%%%%%%%%%%%%%%%%%%%%%%%%%%%%%%%%%%%%%

%%%%%%%%%%%%%%%%% APPENDICES %%%%%%%%%%%%%%%%%%%%%

%\appendix


%%%%%%%%%%%%%%%%%%%%%%%%%%%%%%%%%%%%%%%%%%%%%%%%%%

% Don't change these lines
\bsp	% typesetting comment
\label{lastpage}

\end{document}